\documentclass[twoside,10pt]{article}
\PassOptionsToPackage{square, numbers}{natbib} 
\usepackage{jmlr2e}
\usepackage[utf8]{inputenc} 
\usepackage[T1]{fontenc}    

\usepackage{hyperref}       
\usepackage{url}            
\usepackage{booktabs}       
\usepackage{amsmath}
\usepackage{amssymb}
\usepackage{amsfonts}
\usepackage{amsthm}         
\usepackage{bm}
\usepackage{mathtools}
\usepackage{thmtools}
\usepackage{thm-restate}
\usepackage{nicefrac}       
\usepackage{microtype}      
\usepackage{wrapfig}
\usepackage{multirow}
\usepackage{adjustbox}
\usepackage{natbib}
\usepackage{cleveref}
\usepackage{enumitem}
\usepackage{xcolor}
\usepackage{xspace}
\usepackage{subcaption}
\usepackage{stmaryrd}
\usepackage{float}
\usepackage[version=4]{mhchem}
\usepackage[frak=pxtx]{mathalpha}
\usepackage{mathrsfs}
\usepackage{algorithm}
\usepackage{algorithmic}
\usepackage{diagbox}
\usepackage{comment}
\usepackage{tikz}
\usetikzlibrary{calc,patterns,angles,quotes}

\Crefname{equation}{Eq.}{Eqs.}
\Crefname{figure}{Fig.}{Fig.}
\Crefname{tabular}{Tab.}{Tabs.}
\Crefname{table}{Tab.}{Tabs.}
\Crefname{section}{Sec.}{Sec.}
\Crefname{appendix}{App.}{App.}

\akbcheading
\ShortHeadings{Modelling crypto markets by multi-agent reinforcement learning}{Lussange, Vrizzi, Palminteri, Gutkin}
\finalcopy
\begin{document}

\title{\vspace{-9mm}
Modelling crypto markets by multi-agent reinforcement learning
}

\author{\name Johann Lussange$^{1}$ \email johann.lussange@ens.fr \\
       \name Stefano Vrizzi$^{1}$ \email stefano.vrizzi@ens.fr \\
       \name Stefano Palminteri$^{1,2}$  \email stefano.palminteri@ens.fr \\
       \name Boris Gutkin$^{1,2}$ \email boris.gutkin@ens.fr \\
       \\
      \addr $^{1}$ Laboratoire des Neurosciences Cognitives, INSERM U960, Département des Études Cognitives\\
École Normale Supérieure, 29 rue d'Ulm, 75005, Paris, France. \\
      \addr $^{2}$ Center for Cognition and Decision Making, Department of Psychology, NU University Higher School of Economics, 8 Myasnitskaya st., 101000, Moscow, Russia. }
\maketitle

\begin{abstract}
Building on a previous foundation work~\cite{Lussange2020}, this study introduces a multi-agent reinforcement learning (MARL) model simulating crypto markets, which is calibrated to the Binance's daily closing prices of $153$ cryptocurrencies that were continuously traded between 2018 and 2022. Unlike previous agent-based models (ABM) or multi-agent systems (MAS) which relied on zero-intelligence agents or single autonomous agent methodologies, our approach relies on endowing agents with reinforcement learning (RL) techniques in order to model crypto markets. This integration is designed to emulate, with a bottom-up approach to complexity inference, both individual and collective agents, ensuring robustness in the recent volatile conditions of such markets and during the COVID-19 era. A key feature of our model also lies in the fact that its autonomous agents perform asset price valuation based on two sources of information: the market prices themselves, and the approximation of the crypto assets fundamental values beyond what those market prices are. Our MAS calibration against real market data allows for an accurate emulation of crypto markets microstructure and probing key market behaviors, in both the bearish and bullish regimes of that particular time period. 
\end{abstract}

\section{Introduction}
\label{SectionI}

\paragraph{General problem} Crypto markets, characterized by their high volatility and unpredictability~\cite{segnon2020forecasting}, present a unique challenge for quantitative financial modeling. These markets are not just volatile, they are complex ecosystems influenced by a myriad of factors that are both internal and external to the market~\cite{antonakakis2019cryptocurrency}. Some crypto assets like utility tokens provide users with access to a service such as membership in a network (e.g. decentralized application)~\cite{prat2021fundamental}, others like cryptocurrencies and stable coins serve to facilitate digital transactions and can also be used as a store of value or a unit of account~\cite{wkatorek2021multiscale,lyons2023keeps}, while others like security tokens represent ownership of a real-world asset, such as stocks, bonds, artwork or real estate~\cite{subramanian2020security}. But also, crypto assets rely on very different decentralized technologies~\cite{xiao2020survey}, and each have their own specifics tokenomics~\cite{malinova2023tokenomics,cong2021tokenomics}, with very diverse individual impacts on their valuation. Building on the earlier foundational ideas of distributed ledgers~\cite{nakamoto2009bitcoin,sunyaev2020distributed}, modern crypto assets now also rely on many various protocols~\cite{xu2023survey,gangwal2023survey}. Thus, contrary to more traditional equity markets and their own order book dynamics and transaction orders, these protocols profoundly impact the nature of the crypto market microstructure. For all these reasons, the equity concept of fundamental valuation~\cite{eshraghi2023approaches,benetton2024investors} is notoriously different for crypto assets: unlike stocks, where fundamentals are often linked to tangible economic performance of the company issuing the stocks (e.g. quarterly reports, news of companys' deals), the nature of crypto assets' fundamental values can be more debated and tied to unique features that are proper to crypto assets (e.g. white paper, evolving regulation). Factors such as applicability of the cryptocurrency in real-world transactions, limited supply mechanics (e.g. Bitcoin's fixed supply cap), robustness of the underlying technology, user base and adoption, legal and regulatory developments impacting the asset, etc. are parameters exogeneous to the market that can enter and impact its fundamental value~\cite{beigman2023dynamic,gurdgiev2020herding}. 

\paragraph{Past research} Past research in crypto market modeling that is grounded in traditional financial market theories is thus at risk of being outpaced~\cite{antonakakis2019cryptocurrency,gurdgiev2020herding} by the unique characteristics of crypto markets, such as their extreme volatility, decentralization protocols, and fundamental valuation. Fortunately, recent quantitative methods from agent-based modelling has provided useful tools~\cite{Lussange2018} with a focus on various aspects such as examining the interconnectedness of markets~\cite{Xu2014}, analyzing the effects of market regulation~\cite{Boero2015}, understanding supply and demand dynamics~\cite{Benzaquen2018}, exploring the impact of high-frequency trading~\cite{Wah2013,Aloud2014}, assessing the outcomes of quantitative easing~\cite{Westerhoff2008}, and investigating the influence of external factors~\cite{Gualdi2015}. This domain has seen significant success in replicating characteristic market behaviors, known as \textit{stylised facts}. Stylized facts describe consistent statistical patterns in asset price fluctuations and volatility observed in such diverse markets and time frames that they have been called universal~\cite{Barde2015}. Notable among these observations are those related to the distribution of returns and the phenomenon of volatility clustering, which gained recognition during the 1990s. Pioneering works by Kim and Markowitz~\cite{Kim1989}, Levy, Levy, and Solomon~\cite{Levy1996,Levy1994,Levy1995,Levy1996b,Levy1996c,Levy1997,Levy2000}, Cont and Bouchaud~\cite{Cont2000}, Solomon and Weisbuch~\cite{Solomon2000}, Lux and Marchesi~\cite{Lux1999,Lux2000}, Donangelo and Sneppen~\cite{Donangelo2000,Donangelo2000b,Bak1999,Bak2001}, and Solomon, Levy, and Huang~\cite{Huang2000} significantly contributed to these insights. Notably, it was during this period that ABMs began to emulate these stylized facts. The importance of the universality of these stylized facts stems from the fact that the underlying causes of price movements in different markets can be markedly distinct, encompassing both exogenous and endogenous factors. Consequently, these stylized facts shed light on fundamental financial mechanisms that transcend specific markets. This understanding, in turn, holds considerable value for shaping the architecture of ABMs. From a scientific perspective, the faithful replication of these stylized facts has been a major area of research for approximately the past twenty years~\cite{Lipski2013,Barde2015}. While their precise characteristics have exhibited subtle variations over time and across scholarly works, the most widely recognized accepted stylized facts can be broadly categorized into three intersecting domains: 

\begin{itemize}

\item \textit{Non-Normal Returns}: The distribution of returns notably deviates from the Gaussian distribution, challenging conventional models that treat asset prices as following Brownian random walks~\cite{Potters2001,Plerou1999}. These non-Gaussian distributions exhibit distinctive characteristics, including heavier tails that can be approximated by a power law with exponents typically falling within the range of $[2,4]$. Additionally, these distributions are characterized by negative skewness, often displaying asymmetry in many observed markets, a platykurtic profile, resulting in events that are less centered around the mean~\cite{Cristelli2014,Potters2001,Cont2001,Bouchaud1997,Ding1993,Lobato1998,Vandewalle1997,Mandelbrot1997}, and multifractal $k$-moments, causing the exponent to deviate from a linear relationship with $k$~\cite{Ding1993,Lobato1998,Vandewalle1997,Mandelbrot1997}.

\item \textit{Volatility Clustering}: The volatility of the market exhibits a tendency to cluster or form distinct clusters~\cite{Engle1982}. Consequently, when compared to the average, the likelihood of experiencing high (resp. low) volatility in the near future is notably higher if the recent past witnessed similar high (resp. low) volatility~\cite{Lipski2013,Devries1994,Pagan1996}. Irrespective of whether the next return is positive or negative, this leads to the observation that significant return fluctuations are often followed by similar fluctuations, thus showcasing a form of long-term memory~\cite{Mandelbrot1963} and correlatively, a clustering effect in trading volumes.

\item \textit{Decaying Auto-Correlations}: The auto-correlation function characterizing returns in financial time series data is predominantly near zero for most lag values, with the exception of very short lags. This is attributed to a mean-reverting microstructure mechanism that results in negative auto-correlations~\cite{Cont2001,Cont2005}. This phenomenon aligns with the Efficient Market Hypothesis~\cite{Fama1970,Bera2015}, positing that markets lack memory, implying that the prediction of future prices based on past prices or information is implausible~\cite{Devries1994,Pagan1996}. Nonetheless, it has been observed that certain nonlinear functions of returns, such as squared or absolute returns, exhibit persistent auto-correlations over longer lags~\cite{Cont2005}.

\end{itemize}

Subsequently, ABMs of financial markets have evolved to incorporate increasingly realistic features and are capable of generating more robust scaling experiments. Notably, these simulations hold promise for forecasting actual financial time series data through a reverse-engineering approach~\cite{wiesinger2013reverse,faggini2022toward}. 

\paragraph{New trends} In recent times, ABM in economic studies have gained enhanced realism through the modern breakthrough of machine learning~\cite{vezhnevets2023generative,jumper2021highly} and especially multi-agent learning~\cite{Silver2018,Silver2017}. Reinforcement learning~\cite{Maeda2020,Bjerkoey2019,Rutkauskas2009}, in particular, shares similarities with decision-making processes in the brain~\cite{Momennejad2017,Lefebvre2017,Palminteri2015,Duncan2018}, offering a computational framework to quantitatively analyze agent learning in market price formation. Nonetheless, the integration of reinforcement learning models with real financial data remains an ongoing challenge~\cite{Bjerkoey2019,Rutkauskas2009,Maeda2020}. Although many current models do not fully emulate markets with autonomously learning agents~\cite{Maeda2020,Ganesh2019}, there is a growing interest in applying reinforcement learning to financial ABM~\cite{Ganesh2019,Hu2019,Neuneier1997,Deng2017}, notably crypto markets~\cite{cocco2017using}, as well as order book models~\cite{Spooner2018,Biondo2019,Sirignano2019}. Notably, recent multi-agent reinforcement learning (MARL) models, which consider the collective behavior of multiple market participants, merit consideration~\cite{Lussange2018,Lussange2020, Lussange2022}. Our study builds upon these foundations, seeking to integrate and extend these methodologies, especially in the context of real-world crypto market data from Binance, to provide a more comprehensive understanding of the complexity of crypto markets.

\paragraph{Our contribution} Since Bitcoin's inception in $2009$, the cryptocurrency landscape has expanded significantly, with over $8,000$ digital currencies listed and a cumulative peak value of $2.97$ trillion dollars in $2021$. The rise in crypto exchange volumes reveal the importance of this new highly sought-after alternative asset class. By $2023$, over two hundred exchanges, including giants like Binance with substantial daily trading volumes, have emerged, highlighting the profitability of cryptocurrency trading. In a previous work~\cite{Lussange2020}, we have developed SYMBA (\textit{SYstème Multi-agents Boursier Artificiel}), an innovative MARL stock market simulator, leveraging reinforcement learning for agent autonomy in stock price forecasting and transaction order execution~\cite{Lussange2018}. This model underwent meticulous calibration to align with actual stock market data. Subsequent research~\cite{Lussange2022} explored the evolutionary acquisition of trading strategies by its agents. This model simulates a financial market's microstructure through a bottom-up methodology, utilizing autonomous economic entities—such as investors and institutions—and their financial activities, including trades and holdings. In its equity version, each agent in SYMBA is characterized by two aspects: firstly, a reinforcement learning algorithm for skill enhancement in price prediction and trading activities, allowing them to independently evolve their forecasting and trading proficiency; secondly, their learning orientation is gauged between chartist and fundamentalist, via a weight parameter that is learned by the RL of each agent, depending on the market regime and general state. Agents interact with a centralized order book, submitting individual orders that are then sorted and matched for transactions. Through reinforcement learning, they assess past investment outcomes, adjusting their strategies in response to evolving market conditions. SYMBA not only replicates supply-demand dynamics and other crucial aspects of market behavior like illiquidity and bid-ask spread evolution~\cite{Dodonova2018,Naik2018} but also facilitates comparison of its simulated stock prices and volumes with real financial data. Calibration of the MAS parameters allows for an analysis of the collective impact of agents' learning dynamics. Our model enables a comprehensive examination of financial markets from a micro to macro perspective. In \cite{Lussange2020}, we correlated the overall stock market returns with real-world data, while \cite{Lussange2022} focused on evaluating the agents' trading performance, informed by their evolving strategic learning dynamics. We here apply SYMBA to model crypto markets, via a careful calibration process to Binance's daily close prices of $153$ crypto assets, which were continuously traded from September $27$, $2018$, to September $27$, $2022$, aims to address these challenges.

\paragraph{Structure} The structure of this paper is laid out as such: Section \ref{SectionII} offers a primer on reinforcement learning. Following this, Section \ref{SectionIII} outlines the details of the architecture of our model, SYMBA, outlying its iterative process, agent-specific reinforcement learning algorithms for crypto asset forecasting and trading, and the mechanics of its distributed order book featuring a double auction limit orders procedure. We have made the entire SYMBA code publicly accessible on GitHub for community benefit~\cite{SYMBA_cryptogithub}. The subsequent Section \ref{SectionIV} show its calibration to Binance crypto data, and the optimization procedure of the model hyperparameters, with Section \ref{SectionV} devoted to the results of this calibration. The paper concludes in Section \ref{SectionVI} with a discussion of our results, their implications for cryptocurrency trading and analysis, and potential directions for future research.

\section{Reinforcement learning}
\label{SectionII}

\begin{figure}[!htbp]
\begin{centering}
\includegraphics[scale=0.08]{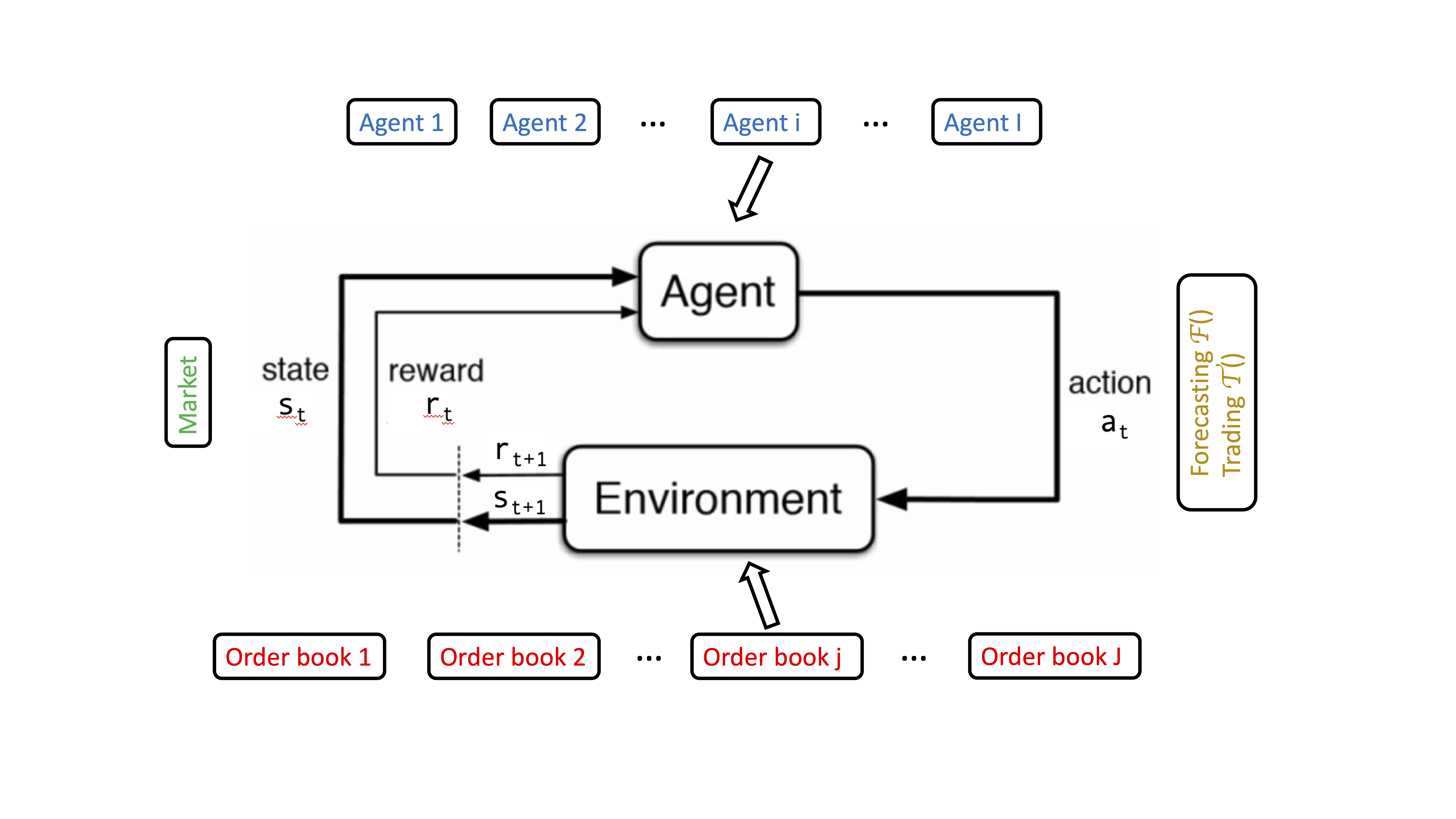}
\caption{\label{F1} Classical algorithmic procedure of a reinforcement learning agent at time step $t$ in the context of SYMBA described in Section \ref{SectionIII}. In a given state $s_t$ of its environment (i.e. the market), a given agent $i$ selects one of its actions $a_t$ (from its forecasting or trading algorithm) with respect to the market order book of a given asset $j$, thus yielding an associated given reward $r_{t+1}$ and new state of the environment $s_{t+1}$.}
\end{centering}
\end{figure}

This section offers a concise overview of the core principles of reinforcement learning, a distinct paradigm within machine learning characterized by its focus on learning through so-called rewards (or reinforcements). For an extensive exploration of this topic, interested readers are directed to key resources such as \cite{SuttonBarto,Wiering2012,Csaba2010}.

\vspace{1mm}

\paragraph{States, actions, rewards}: In the realm of reinforcement learning, three parameters are essential: the set of states $s \in \mathcal S$, depicting the various environments in which the agent may operates without exerting control; the set of actions $a \in \mathcal A$, representing the actions the agent can potentially select in its environment or state $s$; and the reward $r \in \mathbb R$ eventually received by each agent after it selected a given action in a state of its environment $s$ at time $t$. Notably, reinforcement learning assumes Markovian state signals. A state signal is considered Markovian, or possesses the Markov property, if and only if $\forall s', r'$ and prior histories $s_t, a_t, r_t, s_{t-1}, a_{t-1}, r_{t-1}, \ldots, s_1, a_1, r_1, s_0, a_0$, it satisfies:
\begin{equation}
Pr ( s_{t+1} = s', r_{t+1} = r' | s_t, a_t ) \
= Pr ( s_{t+1} = s', r_{t+1} = r' | s_t, a_t, r_t, \ldots, r_1, s_0, a_0 )
\end{equation}

\noindent Following the notation in \cite{SuttonBarto}, $Pr \{X=x\}$ signifies the likelihood of a random variable $X$ assuming the value $x$.

\paragraph{Policy and value functions} To achieve its ultimate objective of maximizing cumulative rewards over time, an intelligent agent must learn and acquire optimal behavior within its environment. This optimal behavior involves determining the most advantageous actions $a$, to undertake in response to specific states of its environment $s$. The function used by the reinforcement learning agent to learn this behavior is called the \textit{policy}, which essentially captures the agent's probabilities of selecting each action $a$ given a particular state $s$. Mathematically, this policy is denoted as:

\begin{equation}
\pi(s,a) = Pr(a | s).
\end{equation}

Initially, this policy begins with equiprobable action selections. However, it evolves over time through exploration by the agent of its environment, ultimately converging towards an optimal policy, denoted as $\pi^{\ast}(s,a)$. In order to do this, at each time step $t$, the agent essentially practices trial and error, selecting a new action $a_t$ while in state $s_t$, and then subsequently estimating the associated rewards. These rewards play a central role in updating the probabilities of each state-action pair within $\pi(s,a)$. Nonetheless, an important feature of reinforcement learning theory is that these rewards aren't always immediate; they can be delayed over time. In the realm of reinforcement learning, this temporal delay in rewards is often addressed using a real parameter known as the \textit{discount factor}, denoted as $\gamma \in (0, 1]$, and introducing the notion of delayed rewards. Here, the agent's policy updates are driven not by the immediate rewards but by the concept of \textit{returns}, denoted as $R_t$. Returns represent the cumulative sum of rewards, factoring in time-based discounting as such over an episodic task of $T \in \mathbb N$ time steps:

\begin{equation}
R_t = \sum_{k=0}^{T} \gamma^k r_{t+k+1}.
\end{equation}

By setting $\gamma<1$, the agent assigns progressively less significance to rewards that are farther into the future, thereby emphasizing the importance of immediate rewards in the learning process. Beyond temporal considerations, it is crucial to recognize that rewards are inherently stochastic or statistical in nature. To assess the magnitude of these returns, the reinforcement learning agent relies on a probabilistic approach by working with the \textit{expectations} of these returns. This idea, coupled with the discount factor, appears in the so-called \textit{value function} used by the agent to evaluate the effectiveness of its action selections. In the model-based version of reinforcement learning, this value function is the \textit{state-value function}, denoted as $V(s)$, and calculates the expected return $R_t$ when the agent is in a specific state $s_t=s$:

\begin{equation}
V(s) = \mathbb{E}[R_t | s_t=s].
\end{equation}

It is itself intimately linked to two fundamental components: the transition probability $\mathcal{P}{ss'}^a=Pr(s_{t+1}=s' | s_t = s, a_t=a)$, representing the likelihood of transitioning from state $s$ to state $s'$ given action $a$, and the expected value $\mathcal{R}{ss'}^a=\mathbb E (r_{t+1} | s_t = s, a_t=a, s_{t+1}=s')$, which represents the expected reward when transitioning from state $s$ to $s'$ under the influence of action $a$. 

In the model-free version of reinforcement learning, the value function used by the agent is the \textit{action-value function}, denoted as $Q(s,a)$. This function quantifies the expected return, considering a specific action $a_t=a$ taken in state $s_t=s$. Formally, it can be expressed as:

\begin{equation}
Q(s,a) = \mathbb{E}\left[R_t | s_t=s, a_t=a\right].
\end{equation}

In summary, these three functions $\pi(s,a)$, $V(s)$, and $Q(s,a)$ are instrumental in shaping the agent's decision-making process, enabling it to learning an optimal policy from its environment effectively, optimize its actions, and adapt to the complex interplay of delayed rewards and stochastic outcomes inherent to reinforcement learning. 

\paragraph{Policy-based vs. value-based}

The convergence of $\pi(s,a)$ to the optimal policy $\pi^{\ast}(s,a)$ is hence the main goal of each reinforcement learning. In order to do this, there are two main families of reinforcement learning approaches. In \textit{policy-based} reinforcement learning, the agent directly updates the probabilities associated with each state-action pair within its policy function. The exploration of the policy space can take place through both "gradient-based" and "gradient-free" techniques. In \textit{value-based} reinforcement learning (such as model-based and model-free reinforcement learning), the value function is first evaluated according to the returns resulting from action selection, and then the policy is updated. In model-based reinforcement learning, the agent first updates the transition probability $\mathcal{P}{ss'}^a$ and the expected value $\mathcal{R}{ss'}^a$, and then derives its value function so as to update its policy. In model-free reinforcement learning, the value function ($V(s)$ or $Q(s,a)$) is directly estimated in order to update the policy function. Model-based reinforcement learning thus relies on the agent converging to the optimal policy by first building a model of its environment (functions $\mathcal{P}{ss'}^a$ and $\mathcal{R}{ss'}^a$) and then deriving the associated value function ($V(s)$ or $Q(s,a)$). The model-free approach offers the advantage of converging to this optimal policy by exploring state-action pairs independently of having a model, streamlining the learning process. Value-based methods rely on the so-called Great Policy Iteration theorem, which consists in updating the value function after the agent received its return, in order to refine the policy, which in turn is used later on at action selection to better evaluate the value function and hence the policy, in a positive feedback loop unto convergence to the optimal policy. Three major different algorithms are used to that end: \textit{Dynamic Programming} (DP), \textit{Monte Carlo} (MC), and \textit{Temporal Difference} (DP) learning. 

\paragraph{Exploration vs. exploitation} Reinforcement learning shares its foundational principles with biology, where optimal behavior is not directly learned but rather emerges through trial and error within the context of predefined rewards or reinforcements. This fundamental process is referred to as the "exploration vs. exploitation" dilemma, where the agent faces a crucial decision over time: should it persistently choose actions that yield the highest rewards (exploitation), or should it explore new actions in the hopes of discovering even better rewards (exploration)? Several methods have been devised to address this exploration-exploitation trade-off. These include the \textit{$\epsilon$-greedy method}, where a probability $\epsilon \ll 1$ is chosen to either explore a random action or exploit the best action at each time step (with $\epsilon$ close to 0 indicating a strong preference for exploitation), the \textit{softmax method}, which ranks actions from best to worst according to a temperature parameter for exploration, and the \textit{pursuit method}, which continually selects the action that is currently deemed the best for exploitation. Balancing exploration and exploitation is a challenge for the agent, as it cannot definitively know whether more exploration will lead to a better policy or if it will yield unfavorable outcomes.

\paragraph{Temporal credit assignment} Moreover, there are instances where a sequence of actions must be undertaken over extended durations before the agent can attain the desired reward. Consequently, reinforcement learning deals with the notion of ``delayed reward.'' In certain applications of reinforcement learning, we hence encounter the accumulation of discounted rewards over time, represented by the return $R_t$ as we saw in the context of state-value $V(s)$ and action-value $Q(s, a)$ functions. This comes with the latter challenge of assigning credit to state-action pairs at each time step, a problem known as \textit{temporal credit assignment}.

\paragraph{Curse of dimensionality} A third crucial feature of reinforcement learning relates to what is often referred to as the \textit{curse of dimensionality}. This issue stems from the fact that in practical applications, the number of possible state-action pairs that an agent must explore rapidly becomes computationally intractable. Consequently, with each reinforcement learning application comes the question to define states and actions in such a way that most state-action pairs can practically be explored by the agent. 

\paragraph{Recent research trends} These considerations regarding exploration versus exploitation, delayed rewards, and dimensionality are central features of reinforcement learning and most active areas of research in recent years have tackled these issues. Notably, \textit{policy gradient methods} seek to refine the control policy iteratively by descending along the gradient of expected returns~\cite{Sutton2000,Silver2014}. Also, the \textit{Q-learning}~\cite{Watkins1992}, itself part of the model-free TD-learning paradigm, has been one of the first major breakthroughs of reinforcement learning, by significantly reducing the number of states that require exploration, focusing solely on pairs $(s,a)$. This issue is crucial when reinforcement learning must deal with large-scale Markov decision processes, making the exploration and convergence towards an optimal policy $\pi^{\ast}(s,a)$ more difficult. This challenge of dimensionality underscores the problem of function approximation, which has led to the fusion of reinforcement learning with artificial neural networks, with the \textit{end-to-end} or \textit{deep reinforcement learning} paradigm~\cite{Silver2016}. Additionally, ongoing research addresses related issues such as \textit{partially observable MDP models}~\cite{Ross2011,Katt2017} and \textit{adversarial reinforcement learning}~\cite{Pinto2017}. These aim to model the noise and uncertainties inherent in state representation by introducing adversarial agents that apply specific perturbations to the system.

A more recent research area is \textit{meta-reinforcement learning}~\cite{Wang2018}, which strives to achieve rapid generalization over different tasks and across varying time scales, a notion closely aligned with \textit{zero-shot and few-shot reinforcement learning}REF. Concurrently, \textit{transfer reinforcement learning}REF aims to transfer the knowledge gained from one task to improve performance on another, while \textit{imitation reinforcement learning}REF focuses on learning from observing other agents. Also, \textit{actor-critic methods} introduce a dual-role paradigm, where an actor shapes the agent's behavior while a critic evaluates the agent's actions~\cite{Grondman2012}. 

Parallel to these developments, multi-agent and \textit{self-play reinforcement learning}~\cite{Heinrich2017} involve agents learning policies by competing against or collaborating with other learning agents. In this domain, \textit{multitask reinforcement learning}REF explores the synergy among multiple tasks to enhance individual task performance. It intersects with \textit{asynchronous reinforcement learning}~\cite{Mnih2016}, which works through parallel agent instances that share a model, diversifying the data collection process. 

\textit{Modular reinforcement learning}~\cite{Andreas2017} decomposes complex tasks into manageable subtasks, crafting individual policies for each, and then reconstructing them into a cohesive policy framework. In this domain, \textit{Monte Carlo Tree Search} (MCTS) reinforcement learning~\cite{Silver2016} adopts a different strategy by employing tree search and random sampling of the search space to determine optimal actions. 

\textit{Lifelong reinforcement learning}~\cite{Tessler2016} deals with the challenge of mastering a multitude of sequential tasks over an agent's lifetime. \textit{Hierarchical reinforcement learning}~\cite{Bhatnagara2006} groups agent actions into higher-level, more abstract tasks, thereby simplifying decision-making. Finally, \textit{hybrid reinforcement learning}REF, often referred to as ``human-in-the-loop'' reinforcement learning, relies on the role of human intervention in improving the learning process, exemplified in domains like intelligent driving. 

This challenge involves the task of specifying and defining the rewards and returns for an agent. Beyond the previously mentioned approach in hierarchical reinforcement learning, research is addressing this issue through \textit{reward shaping}~\cite{Ng1999}, which includes background knowledge related to sub-rewards to enhance convergence rates. Additionally, there is the concept of \textit{inverse reinforcement learning}~\cite{Abbeel2010}, which extracts the reward function from the observation of optimal behavior exhibited by another agent. Furthermore, we can also highlight \textit{homeostatic reinforcement learning}~\cite{Keramati2014,Keramati2011}, which defines the reward within a manifold of numerous sub-rewards, and a state-dependent approach for defining the agent's rewards~\cite{Bavard2018}.

\section{Model and data}
\label{SectionIII}

\begin{figure}[!htbp]
\begin{centering}
\includegraphics[scale=0.075]{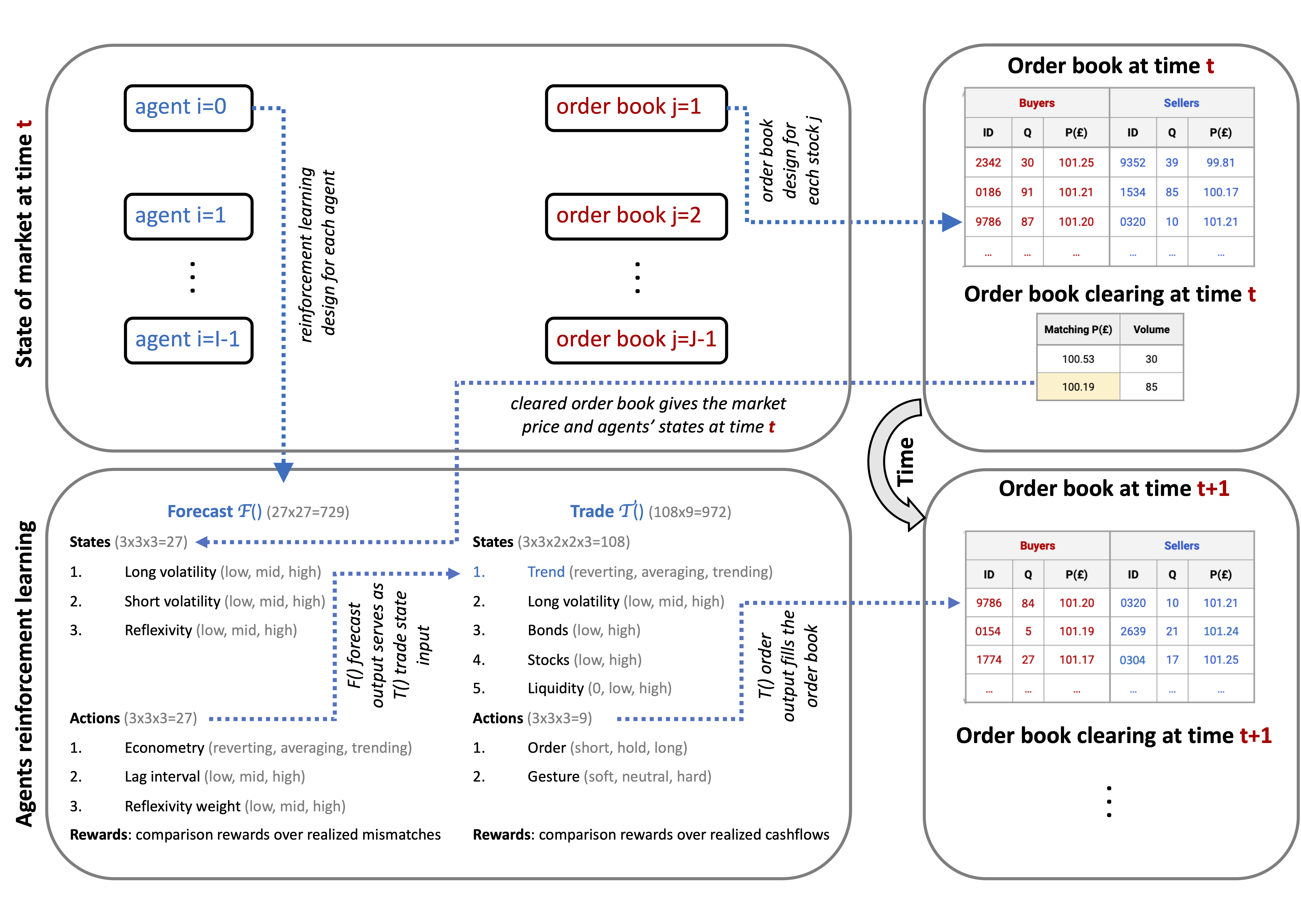}
\caption{\label{F2} Comprehensive Schematic of the SYMBA crypto Market Simulator and Its Operational Dynamics. This figure presents an integrated view of the SYMBA simulator, emphasizing the dual-level interaction within the simulated financial market. At the core of the system, individual agents (bottom-left) utilize two distinct reinforcement learning algorithms, $\mathcal{F}^{i}$ for forecasting and $\mathcal{T}^{i}$ for trading, to independently formulate and execute trading strategies at each simulation step. These strategies are then aggregated at the market level through a centralized double-auction order book (top-right). The order book directs market dynamics by matching buy and sell orders from different agents, effectively determining market prices and volumes (bottom-right). This figure illustrates the iterative loop of agent decision-making and market adjustment (top-left), which collectively shapes the emergent macroscopic market behavior. By simulating the interplay between individual agent strategies and market-level effects, SYMBA provides insights into how individual behaviors and collective market responses yield in a complex financial ecosystem.}
\end{centering}
\end{figure}

In this section, we provide a detailed exposition of the architecture of our crypto market MAS simulator, along with an in-depth discussion of the design principles governing its autonomous agents.

\subsection{General architecture}

As previously said at the end of Section \ref{SectionI}, our simulation relies on several central parameters, namely: the total number of agents denoted $I$, the quantity of assets traded $J$, and the duration of the simulation in discrete time steps $T$. It is worth noting that each time step corresponds to a single trading day, or actually a trading day and night, since contrary to stock markets, crypto markets like Binance trade $24/7$. Thus a year encompasses $T_{y}=365$ trading days, a month spans $T_{m}=30$ trading days, and a week encompasses $T_{w}=7$ trading days. Our analysis typically involves the examination of statistical properties derived from a series of $S$ simulations. Furthermore, our model incorporates critical financial factors such as transaction costs, as network fees $b$ for each crypto asset trade (which can largely vary from exchange to exchange, and from asset to asset), an annual risk-free interest rate $R$ pertaining to agents' risk-free assets, and a general APR or Annual Percentage Rate $D$ (e.g. linked to staking in Proof of Stake consensus protocols for example). For the purpose of simplification and model tractability, we assume these financial parameters to be uniform across simulated crypto assets, with network fees set at $b=0.1\%$, the risk-free rate at $R=1\%$, and the APR at $D=3\%$. Each simulation cycle at time $t$ comprises four primary steps, each outlined below:
\vspace{2mm}

i- \textit{Initialization of Agent Parameters}: At the begining of the simulation, when $t=0$, a number of $I$ agents are initialized, each with their unique set of parameters. These agents, representing either individual or institutional investors, manage portfolios consisting of both crypto assets and risk-free assets (e.g. bank account or bonds). These parameters are elaborated upon in Section \ref{AAgentInitializationParameters}.

\begin{figure}[!htbp]
\centering
\includegraphics[scale=0.36]{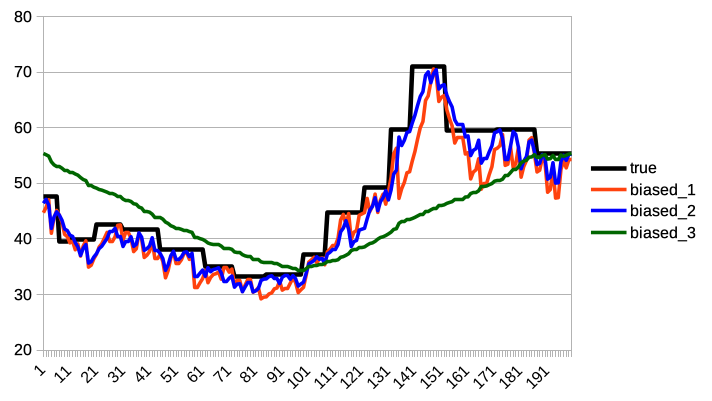}
\caption{\label{Z2} Representation of the trajectories of fundamental values modeled by $\mathcal{T}^{j}(t)$ (depicted as a black line) and their estimated values (denoted as $\mathcal{B}^{i,j}(t)$) by three different agents (displayed as red, blue, and green lines, resp.), over $200$ time steps.}
\end{figure}

\vspace{2mm}

ii- \textit{Initialization of Market Fundamentals}: Following the approach adopted by various models, we set all initial asset prices at $\$ 100$ ($P^{j}(t=0)$). Here we consider the value of the crypto assets in fiat money (i.e. US Dollars) instead of crypto currency (e.g. BTC, i.e. Bitcoin), but this has no impact on the microstructure of the prices and volumes that are generated to emulate crypto markets. Then, $J$ price time series $\mathcal{T}^{j}(t)$ are generated as stochastic jump processes, mirroring the fundamental values of stocks. The term ``fundamental values'' for crypto assets, as used in this context, can be comprehended in relation to the concept of ``fundamental valuation'' found in stock markets. In traditional stock markets, the fundamental value of a share is based on the economic performance and future prospects of the company that issued it. When it comes to various types of crypto assets, such as utility tokens, cryptocurrencies, stablecoins, or digital securities representing real-world assets, their fundamental value is different. It is determined by factors external to their current market price, focusing on the potential technological advancements and economic growth associated with the asset. While there might be debates around this, it is crucial to recognize that investors often approach crypto assets with a mindset similar to that of traditional equities. They believe these assets have an inherent value that is not solely reflected in their market price, encompassing past, present, and future potential. However, it is important to note that this information is not directly accessible to the $I$ agents. Instead, each agent $i$ estimates the value $\mathcal{T}^j(t)$ for asset $j$ using their own cointegration rule denoted as $\mathcal{\kappa}^{i,j} [ \mathcal{T}^{j}(t) ]=\mathcal{B}^{i,j}(t)$. Consequently, the series $\mathcal{B}^{i,j}(t)$ represent the perceived fundamental values of asset $j$ over time $t$ as perceived by agent $i$. As an example, Fig. \ref{Z2} elucidates the notion of cointegration by juxtaposing the modeled fundamental values $\mathcal{T}^{j}(t)$ with their approximations $\mathcal{B}^{i,j}(t)$ as derived by select agents.

From a comprehensive analysis of $S=20$ simulations, we determine that the average annual number of jumps in $\mathcal{T}^j(t)$ above or below $20 \%$ is $1.30$, the average jump amplitude (expressed as $(\mathcal{T}^j(t)-\mathcal{T}^j(t-1))/\mathcal{T}^j(t)$) is $4.97\%$ (with standard deviation of $6.50$), and the average disparity between the biased and actual values (indicated as $(\mathcal{T}^j(t)-\mathcal{B}^{i,j}(t))/\mathcal{T}^j(t)$) amounts to $4.69\%$ (with standard deviation of $1.38$). These insights inform the agents' crypto asset pricing strategies, which incorporate both chartist and fundamental sources of information.

iii- \textit{Autonomous Forecasting and Trading by Agents}: Within this model, each agent independently utilizes two reinforcement learning algorithms for engaging in the market. Further details about these algorithms can be found in Section \ref{AAgentsRL}. The initial algorithm, labeled as $\mathcal{F}^{i}$, focuses on creating an optimal econometric forecasting function. This function takes into consideration the unique characteristics of the market's microstructure and the agent's own fundamental valuation $\mathcal{B}^{i,j}(t)$. The forecast produced by $\mathcal{F}^{i}$ is subsequently utilized by the second reinforcement learning algorithm, $\mathcal{T}^{i}$. The responsibility of $\mathcal{T}^{i}$ is to generate the most suitable limit order for a double auction order book (as explained in~\cite{Mota2016}) at that particular time step. Notice the order books of a Centralized EXchange (CEX) like Binance are not dissimilar from those of more classical markets like stock markets. This process incorporates the forecast as well as various indicators related to the market microstructure and the agent's portfolio. An essential component of this procedure is the filter function $\mathcal{G}^{i}$, which determines the optimal time step for the agent to execute a transaction order.

\vspace{2mm}

iv- \textit{Order Book Population and Processing}: At each time step $t$, a collection of $J$ order books is populated with limit orders from agents for a specific crypto asset $j$. These orders are organized such that buy orders are ranked in descending order of bid prices, and sell orders in ascending order of ask prices, each accompanied by the quantity of crypto assets offered for trade. The clearing of the order book takes place at the same time step $t$. It involves matching buy and sell orders starting from the highest bid and lowest ask prices, progressing until bid prices no longer exceed ask prices. The market price $P^{j}(t+1)$ for asset $j$ in the subsequent time step $t$ is determined by the mid-price at this final matching point. Likewise, the trading volume $V^{j}(t+1)$ is defined as the total quantity of crypto asset $j$ exchanged at time $t$. Additionally, the spread $W^{j}(t+1)$ for asset $j$ at time step $t$ is calculated as the absolute difference between the mean of all bids and asks. Notably, this spread, $W^{j}(t)$, is employed in the agents' crypto asset pricing mechanism, rather than the traditional bid-ask spread, typically defined as the gap between the highest bid and the lowest ask. The pseudo-code of SYMBA's iteration procedure is found in Fig. \ref{FigPseudoCode}.

\begin{figure}
    \begin{verbatim}
Initialize Market parameters at t=0: I, T, P(t=0)...
Initialize Agents parameters at t=0: w, h, B(t=0)...
for (t=1; t<T; t+=1) { // simulation starts
   for (i=0; i<I; i+=1) { // loops over all agents
      for (j=0; j<J j+=1) { // loops over all crypto assets
         Agents[i].F() // each agent RL forecasting
         Agents[i].T() // each agent RL trading
      } // closes j loop
   } // closes i loop
   for (j=0; j<J j+=1) { // loops over all crypto assets
      OrderBook[j].Sort() // sorts offers vs. bids
      OrderBook[j].Clear() // clears transactions, outputs P(t)
   } // closes j loop
   Outputting online statistics...
} // closes t loop
Outputting offline statistics...
    \end{verbatim}
    \caption{Pseudo-code of SYMBA's iteration procedure.}
    \label{FigPseudoCode}
\end{figure}

\subsection{Initialization procedure}
\label{AAgentInitializationParameters}

Our approach incorporates a diverse set of parameters that operate at both the individual agent and the overarching framework levels. As previously outlined in Section \ref{SectionIII}, we utilize continuous and discrete uniform distributions denoted as $\mathcal{U}()$ and $\mathcal{U}{}$, respectively. Additionally, continuous and discrete normal distributions are denoted as $\mathcal{N}()$ and $\mathcal{N}{}$, respectively.
At \textit{step $1$}, each agent $i$ receives the following initial parameter assignments: 

\begin{itemize} 

\item[--] A trading window, denoted as $w^{i}$, is determined by a uniform distribution $\mathcal{U}{ T_w, \tau^{i} }$. This parameter significantly influences the $\mathcal{G}^{i}$ function, which is responsible for determining the optimal timing for crypto assets purchases. 

\item[--] An initial value of risk-free assets denoted as $A_{\text{bonds}}^{i}(t=0)$, following a normal distribution $\mathcal{N}(0,10^4)$. This represents the agent's holdings in bonds or their bank account balance, which fluctuates in response to the agent's decisions to short or long crypto assets.

\item[--] The quantity of crypto assets $Q^{i,j}(t=0)$ for each asset $j$, derived from a discrete positive half-normal distribution $\mathcal{N}^{+} { 0,100 }$. The total value of these assets is calculated as $A_{\text{equity}}^{i}(t=0) = \sum_{j=0}^J Q^{i,j}(t=0) P^{j}(t=0)$. Agents have the option to short sell these crypto assets in the market. Notice the agents cannot buy not sell partial amount of a given crypto asset (e.g. $0.5 ETH$), but only an integer amount thereof. One of our model assumptions is that this integer feature does not change the simulated market microstructure and dynamics significantly, as real crypto market agents often deal with standard fractional values of certain cryptocurrencies for example (cf. millibit or mBTC equal to $10^{-3}$ BTC, satoshi or sat equal to $10^{-8}$ BTC, milliether equal to $10^{-3}$ ETH, gwei equal to $10^{-9}$ ETH, etc.).

\item[--] An investment duration $\tau^{i}$, selected from a uniform distribution $\mathcal{U} {T_w, 6T_m }$. This parameter determines the time frame after which the agent will liquidate its position, with a range spanning from one week to six months in trading days.

\item[--] A memory span $h^{i}$, obtained from a uniform distribution $\mathcal{U} { T_w, T }$. This interval represents the historical data duration considered by the agent during its learning process.

\item[--] A transaction gesture threshold $g^{i}$, derived from a uniform distribution $\mathcal{U}(0.2, 0.8)$. This parameter indicates the agent's willingness to transact at prices either above or below its own asset valuation. The range of this parameter is influenced by the model's gesture scalar $\zeta^i$, as discussed in Table \ref{T1} below. 

\item[--] A reflexivity amplitude parameter $\rho^{i}$, assigned from a uniform distribution $\mathcal{U}(0, 100\%)$. This parameter impacts the agent's approach to price valuation, striking a balance between technical market forecasts and fundamental pricing. It directly influences the action amplitude $\mathcal{F}$ within the first reinforcement learning algorithm.

\item[--] A reinforcement learning rate parameter $\beta^i$, set from a uniform distribution $\mathcal{U}(0.05, 0.20)$. This rate applies to both reinforcement algorithms $\mathcal{F}^{i}$ and $\mathcal{T}^{i}$ and is informed by findings in neuroscience literature~\cite{Palminteri2017,Lefebvre2017,Palminteri2015}.

\item[--] A drawdown threshold $l^{i}$, defined as the year-to-date peak-to-bottom loss in net asset value, drawn from a uniform distribution $\mathcal{U}(40\%, 50\%)$. If the agent's portfolio value falls below this threshold at any given time step $t$, the agent is declared bankrupt and excluded from further market interactions. It's worth noting that this threshold is set higher than typical industry standards to ensure a constant number of agents within our model, even in bankruptcy scenarios. 

\end{itemize}

In this section, we will explain further the various parameters of our model. Some of these parameters are optimized as hyperparameters of the model, such as the drawdown threshold $l^{i}$, which is adjusted using a threshold parameter $\mathcal{L} \in \mathbb N^{+}$). Additionally, the transaction gesture parameter, $g^{i}$, is modified through $\zeta^i \in \mathbb{N}$. Other parameters are derived from existing literature, such as the reinforcement learning rate $\beta^i \in \mathbb{R}^{+}$. Parameters like the reflexivity amplitude parameter, denoted as $\rho^{i}$, are integrated as learned variables in the reinforcement learning process of the agent. Furthermore, certain parameters are preset, including the initial values of agents' bond portfolios $A_{\text{bonds}}^{i}$, equity portfolios $A_{\text{equity}}^{i}$, the investment horizon $\tau^{i}$, and the time intervals $w^{i}$ and $h^{i}$. These components collectively form the foundational framework of our model's architecture.

\vspace{1mm}

\subsection{Agent first RL algorithm} 
\label{AAgentsRL}

In this section, we will explore the details of "step $3$" mentioned earlier, with a focus on the two core reinforcement learning algorithms: $\mathcal{F}^{i}$, responsible for accurate price prediction, and $\mathcal{T}^{i}$, dedicated to effective trading based on those predictions. As previously discussed, each agent $i$ independently runs these algorithms for each crypto asset $j$ at every time step $t$. These algorithms utilize a direct policy search approach, where the probability of each action is directly determined from the policy, bypassing the use of an action-value function as described in the Generalized Policy Iteration theorem~\cite{SuttonBarto}. The action-state pairs for these algorithms consist of $729$ for $\mathcal{F}^{i}$ and $972$ for $\mathcal{T}^{i}$. We define the sets of states $\mathcal{S}$, actions $\mathcal{A}$, and returns $\mathcal{R}$ for both algorithms as follows.
The first algorithm, $\mathcal{F}^{i}$, enables the agent to monitor long-term asset price volatility ($s_0^{\mathcal{F}}$), short-term volatility ($s_1^{\mathcal{F}}$), and the discrepancy between its fundamental valuation and the current market price ($s_2^{\mathcal{F}}$). Using this information, the agent optimizes its price forecasting over its investment horizon $\tau^{i}$ by evaluating three actions through a direct policy search: adopting a basic econometric forecasting method focused on mean-reverting, averaging, or trend-following ($a_0^{\mathcal{F}}$), selecting the duration of the past interval for forecasting ($a_1^{\mathcal{F}}$), and determining the influence of its own fundamental asset pricing in the combined future price estimate, which incorporates both fundamentalist and chartist perspectives ($a_2^{\mathcal{F}}$).

\vspace{3mm}

\textit{State Space $\mathcal{S}^{\mathcal{F}}$}: The $\mathcal{F}^{i}$ algorithm operates within a state space denoted as $\mathcal{S}^{\mathcal{F}}$, which comprises $27$ dimensions. Each component of this state space can assume values of $0$, $1$, or $2$. The agents calculate the variances $\sigma^2_{L}$ and $\sigma^2_{S}$ of crypto asset prices $P^{j}(t)$ over specific time frames. 

\begin{itemize} 

\item[--] $\sigma^2_{L}$ represents long-term volatility and is evaluated and stored as a time series. It is sorted in ascending order and truncated to match the agent's memory span $h^{i}$. The percentile ranking of $\sigma^2_{L}$ at time $t$ determines $s_0^{\mathcal{F}}$, classifying it into three categories: $s_0^{\mathcal{F}}=0$ if it falls below $25\%$, $s_0^{\mathcal{F}}=2$ if it exceeds $75\%$, and $s_0^{\mathcal{F}}=1$ otherwise. 

\item[--] Similarly, $\sigma^2_{S}$, indicating short-term volatility, is processed and categorized in the same manner as $s_1^{\mathcal{F}}$, providing insights into the short-term market dynamics of asset $j$. 

\item[--] The deviation between the market price and the agent's fundamental valuation is measured by averaging the relative difference $\lvert P^{j}(t)-\mathcal{B}^{i,j}(t) \rvert / P^{j}(t)$ over a specific interval $[t-3 \tau^{i}, t]$. This measurement sets $s_2^{\mathcal{F}}=0$ if it is below $10\%$, $s_2^{\mathcal{F}}=2$ if it exceeds $30\%$, and $s_2^{\mathcal{F}}=1$ otherwise. 

\end{itemize}

\vspace{3mm}
\textit{Actions $\mathcal{A}^{\mathcal{F}}$}: Within the framework of the reinforcement learning algorithm $\mathcal{F}^{i}$, we contemplate an action $a^{\mathcal{F}}$, which is a constituent of a set $\mathcal{A}^{\mathcal{F}}={ a_0^{\mathcal{F}}, a_1^{\mathcal{F}}, a_2^{\mathcal{F}} }$, encompassing 27 possible states. Each action, specifically $a_0^{\mathcal{F}}, a_1^{\mathcal{F}}, a_2^{\mathcal{F}}$, holds the capability to independently adopt one of the values $0$, $1$, or $2$. The determination of these actions is subject to a direct policy search, as elaborated subsequently, contingent upon whether the agent finds itself in a phase of exploration or exploitation. Initially, each agent computes two distinct mean values, $\langle P^{j}{[t-2T, t -T]}(t) \rangle$ and $\langle P^{j}{[t-T, t]}(t) \rangle$, derived from the historical asset prices. Here, $T$ is explicitly defined as $T=(1+a_1^{\mathcal{F}}) \tau^{i}/2$. Subsequently, the econometric mechanism calculates:
\begin{equation} \hat{P}^{i,j}(t)=P^{j}(t)+\langle P^{j}{[t-2T, t -T]}(t) \rangle - \langle P^{j}{[t-T, t]}(t) \rangle \end{equation}
\begin{equation} \hat{P}^{i,j}(t)= \frac{1}{2} \langle P^{j}{[t-2T, t -T]}(t) \rangle + \frac{1}{2} \langle P^{j}{[t-T, t]}(t) \rangle \end{equation}
\begin{equation} \hat{P}^{i,j}(t)=P^{j}(t)-\langle P^{j}{[t-2T, t -T]}(t) \rangle + \langle P^{j}{[t-T, t]}(t) \rangle \end{equation}
\noindent These equations apply for $a_0^{\mathcal{F}}=0, 1, 2$, respectively. These actions correspond to strategies encompassing mean-reversion, utilization of moving averages, and tracking trends. Consequently, actions $a_0^{\mathcal{F}}$ and $a_1^{\mathcal{F}}$ are intimately associated with the realm of technical analysis, with $a_0^{\mathcal{F}}$ dictating the selection of the econometric forecasting approach and $a_1^{\mathcal{F}}$ defining the duration of these forecasts. The third action, $a_2^{\mathcal{F}}$, associate the chosen technical forecast $\hat{P}^{i,j}(t)$ with the agent's fundamental valuation $\mathcal{B}^{i,j}(t)$, giving rise to the agent's projection:

\begin{equation} 
H^{i,j}(t)=\alpha \hat{P}^{i,j}(t) + (1-\alpha) \mathcal{B}^{i,j}(t) 
\end{equation}

\noindent In this equation, $\alpha \in \mathbb{R}$ is selected based on the agent's reflexivity $\rho^{i}$. If $\rho^{i} \leq 50\%$, then $\alpha$ assumes the values $0, \rho^{i}, 2\rho^{i}$ for $a_2^{\mathcal{F}}=0,1, 2$, respectively. Conversely, if $\rho^{i} > 50\%$, $\alpha$ takes on the values $2 \rho^{i}-1, \rho^{i}, 1$ for $a_2^{\mathcal{F}}=0,1, 2$. Consequently, when $a_2^{\mathcal{F}}=2$, the agent adapts the weight allocated to its chartist versus fundamentalist valuation methods.

\vspace{3mm}

\textit{Returns $\mathcal{R}^{\mathcal{F}}$}: Subsequently, the reinforcement learning algorithm $\mathcal{F}^{i}$ computes the percentage divergence between the agent's previous asset price prediction $H^{i,j}(t- \tau^{i})$ made $\tau^{i}$ time steps earlier, and the present actual price $P^{j}(t)$:

\begin{equation} 
\frac{\lvert H^{i,j}(t- \tau^{i}) - P^{j}(t) \rvert }{ P^{j}(t)} 
\end{equation}

\noindent This value is logged at each time step, organized in ascending order, and sustained for a duration equivalent to the agent's memory interval $h^{i}$. The percentile rank of this value at time step $t$ is employed to assign a discrete return value $r^{\mathcal{F}}$ from the set $\mathcal{R}^{\mathcal{F}}={ 4,2,1,-1,-2,-4 }$, corresponding to the intervals $[ 0\%, 5\%($, $[ 5\%, 25\%($, $[ 25\%, 50\%($, $[ 50\%, 75\%($, $[ 75\%, 95\%($, $[ 95\%, 100\%]$, respectively.

\vspace{3mm}

\textit{Policy $\mathcal{\pi}^{\mathcal{F}}$}: The reinforcement learning mechanism undergoes regular policy refinement $\pi_t^{\mathcal{F}}(s^{\mathcal{F}}{t- \tau^{i}},a^{\mathcal{F}}{t- \tau^{i}})$ at each time step $t$. The adjustment of this policy is influenced by the learning rate $\beta$. To enhance the likelihood of selecting the optimal action $a^{\mathcal{F}\star}$ in the state $s^{\mathcal{F}}$, a set of equations is iteratively applied $\lvert r^{\mathcal{F}} \rvert$ times. The goal is to increase the probability of this optimal action relative to all other possible actions, denoted $\forall a^{\mathcal{F}} \neq a^{\mathcal{F} \star}$ as:

\begin{equation} 
\pi^{\mathcal{F}}_{t+1} (s^{\mathcal{F}}, a^{\mathcal{F} \star}) = \pi^{\mathcal{F}}_t (s^{\mathcal{F}},a^{\mathcal{F} \star}) + \beta \left[ 1 - \pi^{\mathcal{F}}_t (s^{\mathcal{F}},a^{\mathcal{F} \star}) \right] 
\end{equation}

\begin{equation} \pi^{\mathcal{F}}_{t+1} (s^{\mathcal{F}}, a^{\mathcal{F}}) = \pi^{\mathcal{F}}_t (s^{\mathcal{F}},a^{\mathcal{F}}) - \beta \pi^{\mathcal{F}}_t (s^{\mathcal{F}},a^{\mathcal{F}}) 
\end{equation}

Additionally, the algorithm incorporates an off-policy approach at intervals of $\tau^{i}/T_m + 2$. This approach calculates the action that should have ideally been taken by $\mathcal{F}^{i}$ $\tau^{i}$ steps earlier, using current price and forecast accuracy. The policy $\pi^{\mathcal{F}}$ is then updated using the learning rate $\beta$ and applied $\lvert r^{\mathcal{F}} \rvert=4$ times to adapt to the newly identified optimal action.

\subsection{Agent second RL algorithm} 

This secondary approach allows the agent to dynamically assess the evolution of asset prices, as initially determined by the primary algorithm ($s_0^{\mathcal{T}}$). It also evaluates market volatility ($s_1^{\mathcal{T}}$), the status of risk-averse assets ($s_2^{\mathcal{T}}$), the current quantity of crypto assets held ($s_3^{\mathcal{T}}$), and the volume of executed trades ($s_4^{\mathcal{T}}$). Utilizing this collected information, the agent fine-tunes its investment strategies. It makes decisions on whether to hold, buy, or sell crypto assets in specific quantities ($a_0^{\mathcal{T}}$) and determines the transaction price in response to market supply and demand dynamics ($a_1^{\mathcal{T}}$).

\vspace{3mm}

\textit{States $\mathcal{S}^{\mathcal{T}}$}: The agent's decision-making process in algorithm $\mathcal{T}^{i}$ relies on a state $s^{\mathcal{T}}$ within the set $\mathcal{S}^{\mathcal{T}}={ s_0^{\mathcal{T}}, s_1^{\mathcal{T}}, s_2^{\mathcal{T}}, s_3^{\mathcal{T}}, s_4^{\mathcal{T}} }$. This set encompasses a $108$-dimensional space, where $s_0^{\mathcal{T}}$, $s_1^{\mathcal{T}}$, and $s_4^{\mathcal{T}}$ can take on values from the set ${ 0, 1, 2 }$, and $s_2^{\mathcal{T}}$, $s_3^{\mathcal{T}}$ from ${ 0, 1 }$.

\begin{itemize}

\item[--] The agent calculates the ratio $\mu=(H^{i,j}(t)-P^{j}(t))/P^{j}(t)$ and records it in either $\boldsymbol{\mu}{-}$ or $\boldsymbol{\mu}{+}$ time series, based on its negativity or positivity, respectively. These series are sorted in ascending order and capped to match the agent's memory span $h^{i}$. The agent's current percentile value $\overline{\mu}{-}$ in $\boldsymbol{\mu}{-}$ at time $t$ assigns $s_0^{\mathcal{T}}=0$ if it's under $95\%$, and $s_0^{\mathcal{T}}=1$ if not. Likewise, $\overline{\mu}{+}$ in $\boldsymbol{\mu}{+}$ determines $s_0^{\mathcal{T}}$ to be $1$ if under $5\%$, and $s_0^{\mathcal{T}}=2$ otherwise. The state $s_0^{\mathcal{T}}$ thus represents the econometric prediction $\mu$ from the preceding algorithm $\mathcal{F}^{i}$, indicating a decline, stability, or increase in asset $j$ prices in the forthcoming $\tau^{i}$ time steps.

\item[--] The agent records the previously computed variance $\sigma^2_{L}$ of asset prices $P^{j}(t)$ in a time series over the interval $[t-3 \tau^{i}, t]$. This series undergoes sorting and truncation to align with the agent's memory capacity $h^{i}$. At time $t$, the agent's current percentile value determines $s_1^{\mathcal{T}}$ as follows: $s_1^{\mathcal{T}}=0$ if the percentile is below $33\%$, $s_1^{\mathcal{T}}=2$ if it's above $67\%$, and $s_1^{\mathcal{T}}=1$ otherwise. This aids the agent in assessing long-term asset price volatility.

\item[--] The agent sets $s_2^{\mathcal{T}}=0$ if its risk-free asset value $A_{\text{bonds}}^{i}(t)$ falls below $60 \%$ of its initial value $A_{\text{bonds}}^{i}(t=0)$, and $s_2^{\mathcal{T}}=1$ otherwise. This helps the agent monitor its risk-free asset size for crafting appropriate investment strategies.

\item[--] The agent assigns $s_3^{\mathcal{T}}=0$ if the current value of its crypto asset holdings $A_{\text{equity}}^{i}(t)$ is less than $60 \%$ of the starting value $A_{\text{equity}}^{i}(t=0)$, and $s_3^{\mathcal{T}}=1$ otherwise. This process assists the agent in tracking the value of its crypto asset holdings for strategic decision-making.

\item[--] The agent logs trading volumes $V^{j}(t)$ at each time step in a series, which is subsequently sorted in ascending order and truncated to match the agent's memory period $h^{i}$. The current percentile value at time $t$ determines $s_4^{\mathcal{T}}$ as follows: $s_4^{\mathcal{T}}=0$ if $V^{j}(t)=0$, $s_4^{\mathcal{T}}=1$ if below $33\%$, and $s_4^{\mathcal{T}}=2$ otherwise. This information informs the agent about market activity levels and aids in setting appropriate bid or ask prices for transactions. 

\end{itemize}

\vspace{3mm}

\textit{Actions $\mathcal{A}^{\mathcal{T}}$}: Within the context of the reinforcement learning model $\mathcal{T}^{i}$, we introduce a set of actions denoted by $\mathcal{A}^{\mathcal{T}}={ a_0^{\mathcal{T}}, a_1^{\mathcal{T}} }$, where each action $a^{\mathcal{T}}$ belongs to this set. This set of actions is characterized by a dimensionality of $9$. Both actions, $a_0^{\mathcal{T}}$ and $a_1^{\mathcal{T}}$, can take on discrete values from the set ${ 0, 1, 2 }$, and their values are determined through a process of direct policy search, which is elaborated upon below. The representation of action $a_0^{\mathcal{T}}$ serves a dual purpose: it signifies both the quantity of crypto assets and the type of transaction order (sell, hold, or buy) that the agent chooses to place in the order book. In this framework, each agent adheres to a long-only trading strategy, involving the acquisition of crypto assets at a specific price, holding them for a predefined duration, and eventually selling them, ideally at a higher price. The role of action $a_1^{\mathcal{T}}$ is to express the agent’s willingness to be flexible regarding the trading price. These actions depend on the agent’s evaluation of asset $j$’s price, as determined by the initial algorithm $\mathcal{F}^{i}$. The agent's bid price $P^{i,j}{\text{bid}}(t)$ is formulated as follows: 

\begin{equation} 
P^{i,j}{\text{bid}}(t) = \min [H^{i,j}(t), P^{j}(t)] + g^{i}W^{j}(t-1) 
\end{equation}

\begin{equation} 
P^{i,j}_{\text{bid}}(t) = \min [H^{i,j}(t), P^{j}(t)] 
\end{equation}

\begin{equation} 
P^{i,j}_{\text{bid}}(t) = \min [H^{i,j}(t), P^{j}(t)] - g^{i}W^{j}(t-1) 
\end{equation}

\noindent for different values of $a_1^{\mathcal{T}}$ (namely, $0$, $1$, and $2$ respectively), distinct scenarios emerge. The symbol $g^{i}$ signifies the trading gesture of the agent, while $W^{j}(t-1)$ represents the previous time step's market spread for asset $j$. Consequently, the term $\pm g^{i}W^{j}(t-1)$ captures the agent's adaptable response to trading conditions, influenced by factors like $W^{j}(t-1)$ and the trading volumes denoted by $s_4^{\mathcal{T}}$. The agent's ask price, $P^{i,j}_{\text{ask}}(t)$, is formulated as follows:

\begin{equation}
P^{i,j}_{\text{ask}}(t) = \max [H^{i,j}(t), P^{j}(t)] -  g^{i}W^{j}(t-1) 
\end{equation}

\begin{equation}
P^{i,j}_{\text{ask}}(t) = \max [H^{i,j}(t), P^{j}(t)] 
\end{equation}

\begin{equation}
P^{i,j}_{\text{ask}}(t) = \max [H^{i,j}(t), P^{j}(t)] +  g^{i}W^{j}(t-1)
\end{equation}

\noindent considering the scenario where $a_1^{\mathcal{T}}$ takes on the values $0$, $1$, and $2$ respectively. In this context, $Q^{i,j}(t)$ represents the quantity of asset $j$ held by investor $i$ at time $t$. Specifically, when $a_0^{\mathcal{T}}=0$, it signifies that investor $i$ is placing a sell order for their entire asset $j$ holding at the asking price $P^{i,j}_{\text{ask}}(t)$. Conversely, for $a_0^{\mathcal{T}}=1$, investor $i$ opts to maintain their existing position and not execute any transactions. Lastly, when $a_0^{\mathcal{T}}=2$, it signifies a buy order, with the investor acquiring a quantity of crypto asset $j$ determined by the formula $A_{\text{bonds}}^{i}(t)/[P^{i,j}_{\text{ask}}(t)J]$ at the bid price $P^{i,j}_{\text{bid}}(t)$. 

\vspace{3mm}

\textit{Filter Function $\mathcal{G}^{i}$}: The decision-making process of investor $i$ regarding sending a transaction order to the order book at time step $t$ is dictated by the output of the function $\mathcal{G}^{i}$. This function is designed to introduce a delay in order placement to optimize timing. To achieve this, $\mathcal{G}^{i}$ maintains a time series, recording the maximum value of the action-value function $\arg \max_{a} \mathcal{Q}{t}(s,a)$ at each time step, which is then sorted in ascending order. The decision to execute a trade is determined by comparing the current percentile $p_{\mathcal{Q}}(t)$ of this series with the ratio of time elapsed since the last transaction $k^{i,j}(t)$ to the individual trading window $w^{i}$ of the investor. An order is dispatched to the order book only if the condition $p_{\mathcal{Q}}(t)<k^{i,j}(t)/w^{i}$ is met. It is important to note that while $\mathcal{G}^{i}$ governs the initiation of trades, it does not apply to exit strategies, which are executed based on the investor's predetermined investment horizon $\tau^{i}$.

\vspace{3mm}

\textit{Returns $\mathcal{R}^{\mathcal{T}}$}: The algorithm $\mathcal{T}^{i}$ calculates the change in cash flow resulting from the current net asset value of investor $i$'s portfolio, compared to what it would have been if the actions taken $\tau^{i}$ time steps earlier had not occurred. Mathematically, this is expressed as:

\begin{equation}
Q^{i,j}_{\text{OB}}(t-\tau^{i}) [P^{j}(t) - P^{i,j}_{\text{OB}}(t-\tau^{i})]
\end{equation}

In this equation, $Q^{i,j}_{\text{OB}}(t-\tau^{i})$ and $P^{i,j}_{\text{OB}}(t-\tau^{i})$ represent the quantity and price of crypto asset $j$ cleared in the order book at time $t-\tau^{i}$ for investor $i$ and their trading counterpart. These values may differ from the initial values sent by investor $i$ due to partial order fulfillment and the order book's pricing mechanism, which sets the transaction price at the mid-price in conjunction with the counterparty's order price. These values are logged in a time series at each time step, sorted in ascending order, and truncated to maintain a length corresponding to the investor's memory interval $h^{i}$. The percentile of this value at time $t$ determines the discrete return value $r^{\mathcal{T}}$ in the set $\mathcal{R}^{\mathcal{T}}={ 4,2,1,-1,-2,-4 }$, associated with the intervals $[ 0\%, 5\%]$, $[ 5\%, 25\%]$, $[ 25\%, 50\%]$, $[ 50\%, 75\%]$, $[ 75\%, 95\%]$, and $[ 95\%, 100\%]$.

\vspace{3mm}

\textit{Policy Update Mechanism $\mathcal{\pi}^{\mathcal{T}}$}: In the final phase, the reinforcement learning algorithm adjusts its policy $\pi_t^{\mathcal{T}}(s^{\mathcal{T}}_{t - \tau^{i}},a^{\mathcal{T}}_{t - \tau^{i}})$ after every $\tau^{i}$ time steps following each transaction conducted by the agent. This adaptation is driven by the agent's learning rate $\beta$. The following equations are iterated $\lvert r^{\mathcal{T}} \rvert$ times, with the goal of giving priority to a specific action, denoted as $a^{\mathcal{T}\star}$, in state $s^{\mathcal{T}}$. This is achieved by increasing the policy probability associated with $a^{\mathcal{T} \star}$ over other actions, denoted as $\forall a^{\mathcal{T}} \neq a^{\mathcal{T} \star}$:

\begin{equation}
\pi^{\mathcal{T}}_{t+1} (s^{\mathcal{T}}, a^{\mathcal{T} \star}) = \pi^{\mathcal{T}}_t (s^{\mathcal{T}},a^{\mathcal{T} \star}) + \beta [ 1 - \pi^{\mathcal{T}}_t (s^{\mathcal{T}},a^{\mathcal{T} \star}) ]
\end{equation}

\begin{equation}
\pi^{\mathcal{T}}_{t+1} (s^{\mathcal{T}}, a^{\mathcal{T}}) = \pi^{\mathcal{T}}_t (s^{\mathcal{T}},a^{\mathcal{T}}) - \beta \pi^{\mathcal{T}}_t (s^{\mathcal{T}},a^{\mathcal{T}})
\end{equation}

Moreover, the algorithm incorporates an off-policy mechanism every $\tau^{i}/T_m + 2$ time steps. This mechanism calculates the optimal action that $\mathcal{T}^{i}$ should have taken $\tau^{i}$ time steps earlier, taking into account the realized price and forecast accuracy. Subsequently, it updates the policy $\pi^{\mathcal{T}}$ using the agent's learning rate $\beta$. This update process is repeated $\lvert r^{\mathcal{T}} \rvert=4$ times, as the associated action is considered optimal.

It is essential to emphasize that both algorithms $\mathcal{F}$ and $\mathcal{T}$ employ discretized and handcrafted action-state spaces. This choice is motivated by the necessity to conserve computational resources, but also addresses a certain limitation in applying MAS to financial research, which is the substantial computational power requirement. Additionally, the fundamental basis for defining such state and action spaces is rooted in the \textit{Fundamental Theorem of Asset Pricing}~\cite{Delbaen2011}, where present asset prices are estimated from time-discounted future price expectations. In a similar vein, our reinforcement learning framework for the agent comprises a forecasting component $\mathcal{F}^{i}$ and a trading component $\mathcal{T}^{i}$, a design approach reminiscent of recent models such as~\cite{Spooner2018} (see Section \ref{AAgentsRL} for further details).

\subsection{Calibration to real data}
\label{SectionIV}

\paragraph{Model Assumptions}: The SYMBA model is founded on two core assumptions: i- that the behavior of the simulated agents accurately reflects that of real-world investors, and ii- that the transaction limit orders simulated in the order book faithfully represent the dynamics and characteristics of actual crypto market orders. Regarding the former, our approach simplifies the interaction of any agent, regardless of its behavior or strategy, into three distinct possibilities: buying, selling, or holding assets (a long-only strategy). Concerning the latter, it is worth noting that the dynamics of order books have been extensively documented in the literature~\cite{Huang2015}, including for crypto markets~\cite{puljiz2018market,silantyev2019order}, allowing for a rigorous design. 

\vspace{1mm}

\paragraph{Model limitations}: Alongside our fundamental assumptions, we also acknowledge several constraints and consistency challenges that are inherent to all financial MAS: i- Dependency on the generation of virtual fundamentals $\mathcal{T}^{j}(t)$ (notably, wasn't it $\mathcal{B}^{i,j}(t)$ regardless?). ii- The absence of diversification across various asset classes. iii- The lack of diverse trading strategies available on Binance, such as short-selling, leveraging, derivatives, metaorders, market orders, and more. iv- Neglecting intraday and seasonal market effects. v- Disregarding legal and regulatory constraints, which furthermore have been very dynamic over the considered time period of $2018-2022$. While some of these limitations may pose challenges, their impact and importance are intrinsic to nearly all econometric and modeling approaches within the realm of quantitative finance. Furthermore, modeling market activity through a market microstructure derived from a centralized order book that processes transaction orders from multiple trading agents aligns closely with a CEX like Binance, making it empirically relevant.

\vspace{1mm}

\begin{figure}[!htbp]
\begin{centering}
\includegraphics[scale=0.53]{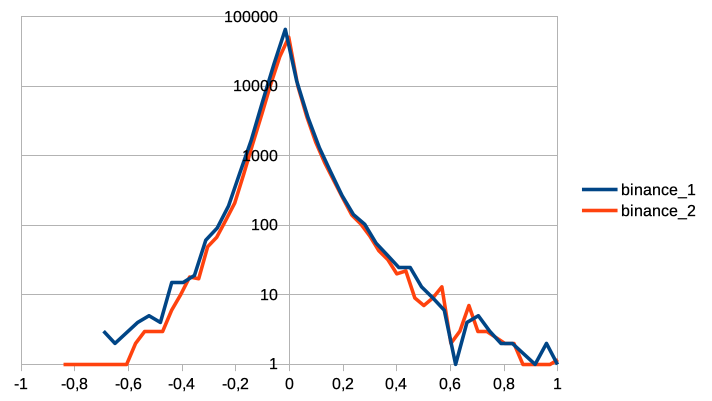}
\caption{\label{AAAA_trunk} Comparative Distribution of Logarithmic Price Returns: The red curve represents the Binance training set, while the blue curve represents the Binance testing set.}
\end{centering}
\end{figure}

\paragraph{Training and testing data}: Our model was calibrated to authentic crypto market data~\footnote{These computations were carried out on a Mac Pro equipped with a 3.5 GHz 6-Core Intel Xeon E5 processor and 16 GB of 1866 MHz DDR memory}. To achieve this, we employed high-quality, industry-standard daily closing prices and trading volumes from CryptoTick, for $153$ crypto assets listed on Binance. This dataset covers the period from September $27$th, $2018$, to September $27$th, $2022$. As mentioned in Section \ref{SectionIII}, these records encompass the date, opening price, highest price, lowest price, closing price, and trading volume for each day. Notably, these figures originate directly from Binance and are not an amalgamation of data from smaller exchanges (i.e., consolidated data). In our analysis of market microstructure, we only included crypto assets that sustained continuous trading during the specified timeframe. Consequently, our initial crypto asset universe was reduced to $153$ stocks. We tuned the MAS hyperparameters using a random sample comprising half of these stocks as a training set. Remarkably, we observed a high level of statistical stability within the training set when compared to the other half, as shown with the logarithmic price returns distributions of these two sets on Fig. \ref{AAAA_trunk}. 

Finally there are three major features which are proper to crypto markets that should here be taken into account, when considering these results: i- There are two main ways to assess the value of a given crypto asset, namely either in fiat money (USD, EUR, GBP, etc.) or in a different crypto currency (BTC, ETH, BNB, etc.). Given the historical central role and major influence played by Bitcoin on all the other crypto assets prices in those markets, studying the microstructure of crypto assets in BTC as we did here, or in USD can be vastly different tasks. ii- Contrary to classical stock markets, trading and transactions in crypto markets run $24/7$. iii- just as Over-The-Counter (OTC) derivatives and other synthetic replications of stocks like Contract for Differences (CFD) proposed by brokers, a vast amount of crypto assets are not traded via a CEX like Binance but Decentralized EXchanges (DEX), with their own decentralized book dynamics. 

\vspace{1mm}

\paragraph{Optimization and hyperparameters}: The hyperparameters subjected to calibration encompass the number of agents ($I$), the agent transaction gesture factor ($\zeta^i \in \mathbb{N}$, which scales the gesture parameter $g^i$ initialized for each agent at $t=0$), the parameter $\nu$ governing the cointegration accuracy of each agent approximating the fundamental time series $\mathcal{T}^j$, and the drawdown threshold (the upper limit of the drawdown, initialized at $t=0$ for each agent). We assessed various combinations of hyperparameters against the training dataset, and the specifics are outlined in Table \ref{T1}. The optimization procedure entailed a maximum of $480$ simulations, each comprising $S=20$ runs to ensure statistical reliability.

\begin{table}[h!]
\caption{Model Hyperparameters and Ranges for Training: Lower Bound (Low), Upper Bound (High), and Increment Step (Step).}
\label{T1}
\centering
\begin{tabular}{ | m{5cm} | m{1cm} | m{1cm} | m{1cm} | } 
\hline
\textbf{Hyperparameter} & \textbf{Low} & \textbf{High} & \textbf{Step} \\ 
\hline
Number of Agents ($I$) & 500 & 5500 & 1000 \\ 
\hline
Gesture Scalar ($\zeta^i$) & 1.0 & 3.0 & 0.5 \\ 
\hline
Cointegration accuracy ($\nu$) & 9 & 12 & 1 \\ 
\hline
Drawdown Threshold ($\mathcal{L}$) & 10 & 90 & 20 \\ 
\hline
\end{tabular}
\end{table}

\vspace{1mm}
\paragraph{Exploring sensitivity}: Throughout the optimization process, we conducted a sensitivity examination to evaluate the model's response to diverse hyperparameter ranges. The objective was to pinpoint regions of non-linearity concerning the alignment with actual data. Notably, we found that augmenting the count of agents ($I$) displayed a linear correlation with the reduction of short-term price fluctuations. Additionally, increasing the values of the gesture scalar ($\zeta^i$) and the cointegration accuracy ($\nu$) exhibited a linear escalation in absolute daily price returns. Remarkably, large drawdown thresholds ($\mathcal{L} > 30\%$) showed minimal impact on agent survivability rates.

\vspace{1mm}


\vspace{1mm}

\begin{figure}[!htbp]
\begin{centering}
\includegraphics[scale=0.4]{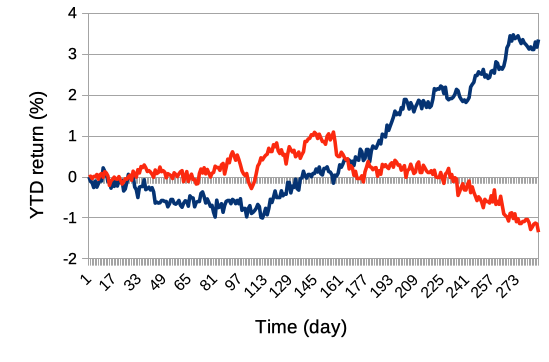}
\includegraphics[scale=0.4]{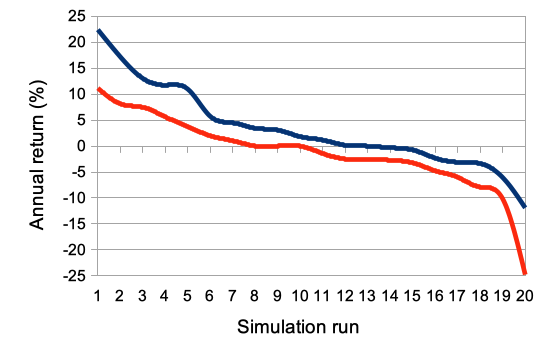}
\caption{\label{AA13} Towards the end of $90\%$ of the total simulation period, our objective is to compare the top-performing $10\%$ of agents in our MAS crypto market simulator, represented by the blue curves, with the top $10\%$ performers in a market simulated with randomly trading noise agents, shown by the red curves. This comparison is based on their performance during the remaining $10\%$ of our overall simulation duration, utilizing averaged equity curves as their year-to-date returns across $20$ simulations (left), and the averaged, sorted annual returns from each of these $20$ simulations (right). These simulations are generated using the following parameters: $I = 500$, $J = 1$, and $T = 2875$.}
\end{centering}
\end{figure}

\paragraph{Comparative model analysis}: The existing literature, as exemplified in~\cite{Gode1993}, studies the substitution of market dynamics for individual rationality. It explores whether markets eliminate irrational participants or whether individuals adapt and learn market rules. Figure \ref{AA13} illustrates learning curves for agents, providing a basis for model comparison, especially when compared to contemporary order book models integrated with reinforcement learning~\cite{Spooner2018}, and the earlier generation of MAS featuring zero-intelligence agents~\cite{Gode1993}, which serve as baseline references.

\vspace{3mm}

\section{Results}
\label{SectionV}

\begin{figure}[!htbp]
\begin{centering}
\includegraphics[scale=0.5]{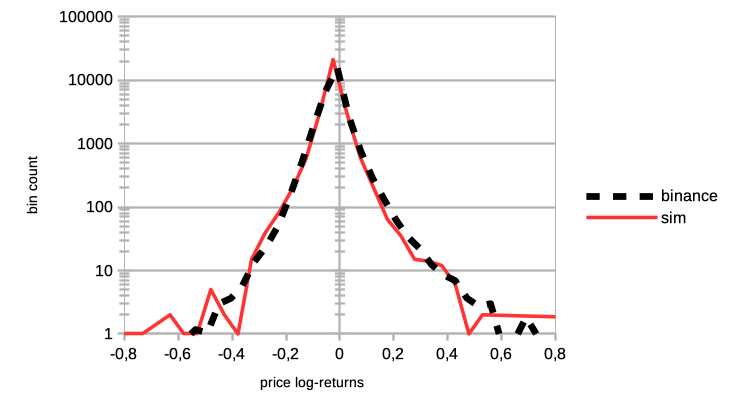}
\caption{\label{R1} Comparative Distribution of Logarithmic Price Returns: The dashed black curve represents real data, while the continuous red curve represents simulated data. These simulations were generated using parameters $I=500$, $J=1$, $T=1453$, and $S=20$.}
\end{centering}
\end{figure}

\begin{figure}[!htbp]
\begin{centering}
\includegraphics[scale=0.5]{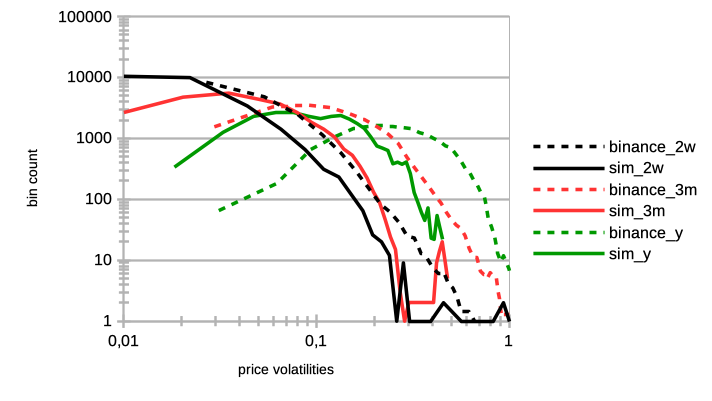}
\caption{\label{R2} Volatility Distribution at Different Time Lags: This figure illustrates the distribution of volatilities, computed at two weeks (black), three months (red), and one year (blue) intervals for both real (dashed curves) and simulated (continuous curves) data. The simulations were generated using parameters $I=500$, $J=1$, $T=1453$, and $S=20$.}
\end{centering}
\end{figure}

\begin{figure}[!htbp]
\begin{centering}
\includegraphics[scale=0.5]{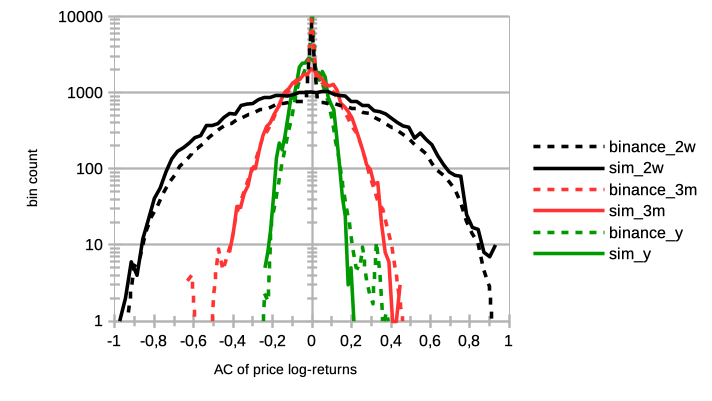}
\caption{\label{R3} Autocorrelations of Logarithmic Price Returns: This figure presents the distribution of autocorrelations of logarithmic returns of prices at each time step $t$ between intervals $[t-\Delta, t]$ and $[t-2\Delta, t-\Delta]$ over lags $\Delta$ of two weeks (black), three months (red), and one year (blue) intervals for both real (dashed curves) and simulated (continuous curves) data. The simulations were generated using parameters $I=500$, $J=1$, $T=1453$, and $S=20$.}
\end{centering}
\end{figure}

\begin{figure}[!htbp]
\begin{centering}
\includegraphics[scale=0.5]{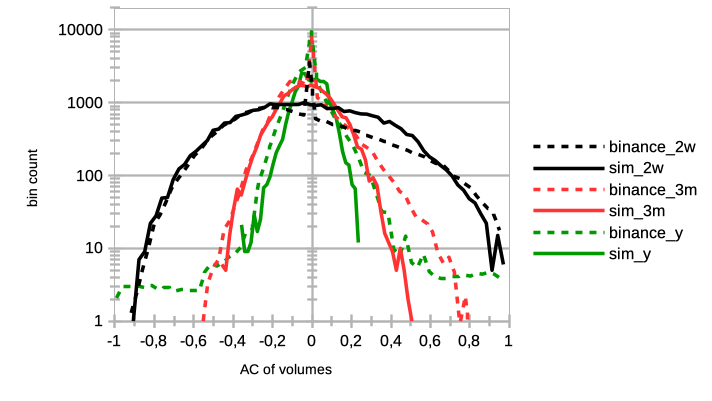}
\caption{\label{R4} Autocorrelations of Trading Volumes: This figure presents the distribution of autocorrelations of trading volumes at each time step $t$ between intervals $[t-\Delta, t]$ and $[t-2\Delta, t-\Delta]$ over lags $\Delta$ of two weeks (black), three months (red), and one year (blue) intervals for both real (dashed curves) and simulated (continuous curves) data. The simulations were generated using parameters $I=500$, $J=1$, $T=1453$, and $S=20$.}
\end{centering}
\end{figure}

\begin{figure}[!htbp]
\begin{centering}
\includegraphics[scale=0.5]{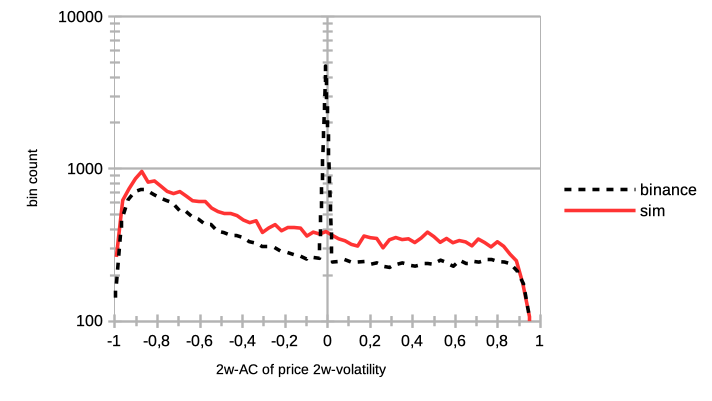 }
\caption{\label{R5} Autocorrelations of Two Weeks-Interval Volatilities: This figure illustrates the distribution of autocorrelations of two weeks-interval volatilities at each time step $t$ between intervals $[t-\Delta, t]$ and $[t-2\Delta, t-\Delta]$ for $\Delta=2T_{w}$, for both real (dashed black curve) and simulated (continuous red curve) data. The simulations were generated using parameters $I=500$, $J=1$, $T=1453$, and $S=20$.}
\end{centering}
\end{figure}

\begin{figure}[!htbp]
\begin{centering}
\includegraphics[scale=0.5]{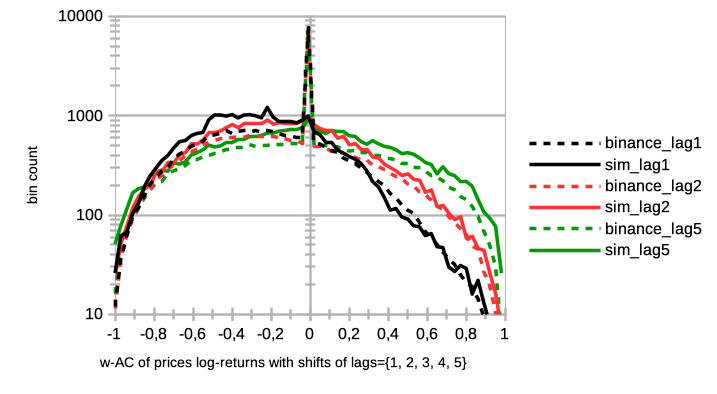}
\caption{\label{R6} Means of Autocorrelations of Logarithmic Price Returns with Various Time Shifts: This figure displays the means of autocorrelations of logarithmic returns of prices at each time step $t$ between intervals $[t-T_w, t]$ and $[t-T_w-\partial, t-\partial]$, for shifts $\partial=[1, 2, 5]$. The data is shown in both real (blue) and simulated (red) scenarios. The simulations were generated using parameters $I=500$, $J=1$, $T=1453$, and $S=20$.}
\end{centering}
\end{figure}

\begin{figure}[!htbp]
\begin{centering}
\includegraphics[scale=0.5]{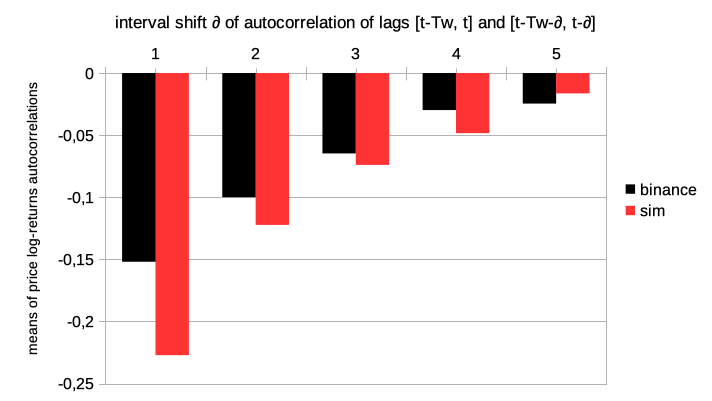}
\caption{\label{R7} Means of Autocorrelations of Logarithmic Price Returns with Extended Time Shifts: This figure demonstrates the means of autocorrelations of logarithmic returns of prices at each time step $t$ between intervals $[t-T_w, t]$ and $[t-T_w-\partial, t-\partial]$, for shifts $\partial=[1, 2, 3, 4, 5]$. The data is presented in both real (blue) and simulated (red) scenarios. The simulations were generated using parameters $I=500$, $J=1$, $T=1453$, and $S=20$.}
\end{centering}
\end{figure}

\begin{figure}[!htbp]
\begin{centering}
\includegraphics[scale=0.5]{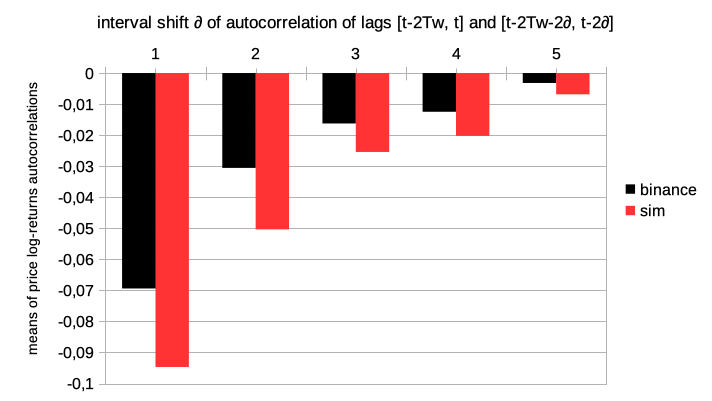}
\caption{\label{R8} Means of Autocorrelations of Logarithmic Price Returns with Extended Time Shifts: This figure demonstrates the means of autocorrelations of logarithmic returns of prices at each time step $t$ between intervals $[t-2T_w, t]$ and $[t-2T_w-2\partial, t-2\partial]$, for shifts $\partial=[1, 2, 3, 4, 5]$. The data is presented in both real (blue) and simulated (red) scenarios. The simulations were generated using parameters $I=500$, $J=1$, $T=1453$, and $S=20$.}
\end{centering}
\end{figure}

In this section, we present a comprehensive set of critical market microstructure indicators relevant to the calibration of our SYMBA model. Figures \ref{R1} through \ref{R8} demonstrate the qualitative agreement in shape between the curves generated by our model and those obtained from actual stock market data. Unless otherwise specified, the following results stem from simulations conducted with $I=500$ agents, $J=1$ traded asset, $T=1453$ time steps per simulation, equivalent to four years of trading days, and a total of $S=20$ simulation runs.

To begin with, Fig. \ref{R1} illustrates the distribution of logarithmic returns of prices $\log [P(t)/P(t-1)]$ for real (dashed black curve) and simulated (continuous red curve) data, displaying a close match between the simulated and real logarithmic price returns. One should notice the limited variability of extreme events in the tails of the distribution, as highlighted on the logarithmic y-axis. In Fig. \ref{R2}, we plot the distributions of price volatilities over different time intervals: two weeks (black), three months (red), and one year (green), for both real (dashed curves) and simulated (continuous curves) data. These volatilities are computed as standard deviations of prices normalized by the price itself, $\sigma/P(t)$. Although the general shapes of the simulated and real data curves are similar, we find a shift or translation in between them. First, we observe that emulating real volatilities at longer time scales is more challenging, likely due to our real data sample covering a unique and exceptional market period during these years, namely the $2020-2022$ crypto market bubble and crash. Secondly, we generally find lower volatility events, at all time scales, for the simulated data than for the Binance data. This may be explained by the fact that a significant portion of cryptocurrency trading was driven during this time period by speculative investors looking for quick returns, notably during the pandemics. This speculative behavior, often based on market sentiment rather than fundamental value of the crypto assets, can lead to rapid price swings. 

We then want to check on the micro-tructure of the price volatility and traded volume, with the clustering features mentioned in Section \ref{SectionI} in mind. In Fig. \ref{R4}, we show the distributions of correlations in the trading volumes between distinct intervals $[t-\Delta, t]$ and $[t-2\Delta, t-\Delta]$ at each time step $t$, considering values of $\Delta$ corresponding to two weeks (black), three months (red), and one year (green). Again, this is for both real (dashed curves) and simulated (continuous curves) data. One can see the great fits of the simulation to real data, highlighting the aforementioned stylized fact of volume clustering. However, one should note the presence of numerous extra zero volume correlations in simulated data wrt. real data, which can be explained similarly as before. Fig. \ref{R5} illustrates the simulation data (shown as the red continuous curve) emulating the actual Binance data (represented as the black dashed curve) wrt. the distribution of price volatility correlations between distinct time intervals $[t-2T_w, t]$ and $[t-4T_w, t-2T_w]$ at each time step $t$. Although we see a strong correspondance in fit shapes, notably in its asymmetry, there is an even larger amount of zero autocorrelation events in simulated data. Apart from the lack of arbitrage efficiency in crypto markets seen before in Fig. \ref{R3}, this surge can be explained by the granularity of the model data, which cannot capture all the intraday events of the real data. Simulated data might thus smooth out certain price variations present in real data, thereby affecting the correlation structure. 

Another key topic of financial quantitative research is that of market memory and efficiency. An important issue for the model' simulated data is hence to check on its emulation of prices autocorrelations. Firstly, Fig. \ref{R3} presents the distributions of correlations in the price logarithmic returns between distinct intervals $[t-\Delta, t]$ and $[t-2\Delta, t-\Delta]$ at each time step $t$, considering values of $\Delta$ corresponding to two weeks (black), three months (red), and one year (green). This analysis is conducted for both real (dashed curves) and simulated (continuous curves) data. Despite the overall good fit, especially regarding the general shape of the distributions, one should note the presence of numerous extra zero autocorrelations in simulated data. We suggest that this difference with real data could be attributed to a still-present lack of arbitrage efficiency in crypto markets, especially during the early period of the $2018-2022$ interval. In Fig. \ref{R6}, the distributions of the means of the correlations in logarithmic price returns at each time step $t$ are presented for simulated data (continuous curves) and actual data (dashed curves), between time intervals $[t-T_w, t]$ and $[t-T_w-\partial, t-\partial]$. This is for shifts $\partial$ ranging from one day (black), to two days (red), to five days (green). Again, the shapes of the simulated curves fit very well the real data, but one can notice the surplus of zero correlations for the simulated data, for reasons akin to the previous point. Fig. \ref{R7} displays the mean correlations of logarithmic returns in prices at each time step $t$ between intervals $[t-T_w, t]$ and $[t-T_w-\partial, t-\partial]$ for shifts $\partial=1, 2, 3, 4, 5$. This comparison is made between real data (in black) and simulated data (in red). Similarly, Figure \ref{R8} presents a close fit analysis for wider intervals of $[t-2T_w, t]$ and $[t-2T_w-2\partial, t-2\partial]$. These statistical insights are crucial for comprehending that our model generates a price microstructure devoid of arbitrage opportunities while displaying the typical decay of market memory through agent learning. In simpler terms, agents learn to exploit short-term causal structures present in historical prices, thereby rending the market more efficient. 

To summarize, the calibration process shows our MARL model ability to replicate the distribution of logarithmic price returns (Fig. \ref{R1}), the distribution of normalized price volatilities at different time scales (Fig. \ref{R2}), and the autocorrelations of trading volumes (Fig. \ref{R4}) and log-price returns across various time scales and intervals (Fig. \ref{R3}, \ref{R5} to \ref{R8}). The stylized facts mentioned in Section \ref{SectionI} of non-gaussian price returns, clustered volatilities and volumes, and decaying autocorrelations are thus re-enacted by SYMBA, and fit the real data. Notably, these autocorrelation metrics play a pivotal role in the calibration procedure as they pertain to the absence of arbitrage opportunities and the market's memory, both being fundamental attributes of financial markets. In simpler terms, beyond the stylized facts, the synthetic data generated by the model should not exhibit discernible price patterns that are more exploitable for trading than those found in real-world data. Additionally, it is worth highlighting that our MARL simulator emulates the dynamics of real stock markets during the very volatile and active regime of crypto markets over the period of $2018$ to $2022$, encompassing periods of economic downturns and growth, as well as the COVID-19 era. 

\vspace{1mm}

That being said, there are several avenues for improving the model's performance and characteristics. Firstly, we should be further addressing the tail distribution fit of long-term price volatilities, as evident in Fig. \ref{R2}. These tails present a formidable challenge to capture, given their association with jump diffusion processes inherent to volatile events affecting very diverse crypto assets (e.g. stable coins, industrial tokens, cryptocurrencies), and as aforementioned, the specific era of COVID-19.Secondly, we should be further investigating the peak in zero autocorrelations observed in real price returns and volatility, showcased in Figs. \ref{R3} and \ref{R4}, and especially Fig. \ref{R5} and \ref{R6}. As said, this phenomenon may be attributed to the simulator's lack of consideration for intraday crypto market activities. It could also potentially be explained by a remaining lack of maturity of crypto markets in terms of market efficiency and no-arbitrage.

\section{Discussion} 
\label{SectionVI}

In this study, we have demonstrated how our MARL model SYMBA can replicate critical aspects of crypto market dynamics, as evidenced through its calibration with actual Binance data from $2018/09/27$ to $2022/09/27$. Our results offer valuable insights into the complexity and behavior of crypto markets, highlighting the model's ability to emulate key market characteristics such as non-normal returns, volume clustering, and decaying price log-returns autocorrelations. These findings underscore the importance of considering the unique attributes of crypto markets, particularly their high volatility, decentralized nature, and diverse factors impacting fundamental valuation, in the financial modeling of crypto assets. The successful calibration of SYMBA to real-world crypto data, despite certain limitations, represents a significant step forward in the understanding and analysis of such markets. It demonstrates the potential of agent-based models, enhanced by reinforcement learning, to capture complex market phenomena that traditional financial models may struggle to represent.

However, our analysis also identified areas for improvement in the model. The observed discrepancies in the tail distribution of long-term price volatilities and the surplus of zero autocorrelations in logarithmic price returns within the simulated data point to the need for further refinement. These issues may stem from the model's current limitations in capturing the very particular market regimes that were those of the $2018-2022$ period, with its specific bubble and crash, together with the effects of the COVID-19 era. Another issue may arise from granularity of our model, and the specific microstructure of crypto markets stemming from the intraday activity, which is actually a $24/7$ trading activity. Addressing these challenges will require enhancing the decision-making processes of market participants, an especially the impact of exogenous factors such as global economic events akin to those aforementioned. Looking ahead, future research should focus on refining the model to better account for the diverse and rapidly evolving nature of crypto assets and market conditions. This includes incorporating more granular data, such as intraday trading activities, and improving the model's ability to simulate market responses to external shocks and policy changes. Additionally, taking into account the ever-increasing impact of new regulations in crypto markets, could further enhance the model's realism.

In conclusion, our study contributes to the growing body of research on crypto markets, offering a novel approach to understanding these complex and dynamic systems. By leveraging the power of agent-based modeling and reinforcement learning, SYMBA provides a framework for future potential applications in crypto markets wrt. risk management, regulatory policy development, and investment strategy optimization. The insights gained from this research not only advance our theoretical understanding of financial markets but also have practical implications for investors, policymakers, and researchers in the field.

\section*{Acknowledgement}

We graciously acknowledge this work to have been supported by the ANR (Agence nationale de la Recherche) CogFinAIgent. Parts of this research were also carried out within the European Union’s Horizon 2020 research and innovation programme under the Marie Skłodowska-Curie grant agreement No $945304$ - Cofund AI4theSciences hosted by PSL$^{\ast}$ University.


\newpage
\nocite{}
\bibliographystyle{unsrtnat}
\bibliography{main}

\renewcommand{\theHsection}{\arabic{section}}

\end{document}